\documentclass[prb,aps,amsmath,preprint]{revtex4}
\usepackage{graphicx}
\usepackage{color}
\usepackage[version=3]{mhchem}
\usepackage{dcolumn}
\usepackage{multirow}
\usepackage{bm}
\usepackage{subfigure}
\usepackage{amssymb}

\begin{document}

\title{Phase stability and lattice dynamics of ammonium azide under hydrostatic compression}
\author{N. Yedukondalu$^1$, G. Vaitheeswaran$^{1*}$, P. Anees$^2$, and M. C. Valsakumar$^3$}
\address{$^1$Advanced Centre of Research in High Energy Materials (ACRHEM),
University of Hyderabad, Prof. C. R. Rao Road, Gachibowli, Hyderabad 500 046, Telangana, India. \\
$^2$ Materials Physics Division, Indira Gandhi Centre for Atomic Research, Kalpakkam 603102, Tamil Nadu, India. \\
$^3$ Department of Physics, IIT Palakkad, Kozhipara, Palakkad 678557, Kerala, India.}
\vspace{5in}
\date{\today}

\begin{abstract}
We have investigated the effect of hydrostatic pressure and temperature on phase stability of hydro-nitrogen solids using dispersion corrected Density Functional Theory calculations. From our total energy calculations, Ammonium Azide (AA) is found to be the thermodynamic ground state of N$_4$H$_4$ compounds in preference to Trans-Tetrazene (TTZ), Hydro-Nitrogen Solid-1 (HNS-1) and HNS-2 phases. We have carried out a detailed study on structure and lattice dynamics of the equilibrium phase (AA). AA undergoes a phase transition to TTZ at around $\sim$ 39-43 GPa followed by TTZ to HNS-1 at around 80-90 GPa under the studied temperature range of 0-650 K. The accelerated and decelerated compression of $a$ and $c$ lattice constants suggest that the ambient phase of AA transforms to a tetragonal phase and then to a low symmetry structure with less anisotropy up on further compression. We have noticed that the angle made by Type-II azides with $c$-axis shows a rapid decrease and reaches a minimum value at 12 GPa, and thereafter increases up to 50 GPa. Softening of the shear elastic moduli is suggestive of a mechanical instability of AA under high pressure. In addition, we have also performed density functional perturbation theory calculations to obtain the vibrational spectrum of AA at ambient as well as at high pressures. Further, we have made a complete assignment of all the vibrational modes which is in good agreement with the experimental observations at ambient pressure. Also the calculated pressure dependent IR spectra show that the N-H stretching frequencies undergo a red and blue-shift corresponding to strengthening and weakening of hydrogen bonding, respectively below and above 4 GPa. Intensity of the N-H asymmetric stretching mode B$_{2u}$ is found to diminish gradually and the weak coupling between NH$_4$ and N$_3$ ions makes B$_{1u}$ and B$_{3u}$ modes to be degenerate with progression of pressure up to 4 GPa which causes weakening of hydrogen bonding and these effects may lead to a structural phase transition in AA around 4 GPa. Furthermore, we have also calculated the phonon dispersion curves at 0 and 6 GPa and no soft phonon mode is observed under high pressure.
\end{abstract}

\maketitle
\section {Introduction}
Ammonium Azide (AA) has been attracting considerable interest as a precursor for synthesis of polymeric nitrogen due to its high nitrogen content of 93.3$\%$ by weight and also the fact that it gives out environment friendly decomposition products.\cite{wu} It also exhibits extensive hydrogen bonding between ammonium cation and terminal nitrogen atoms of azide anions.\cite{prince,amorim} Unlike the alkali, alkaline-earth and transition metal azides which are either ionic or covalent,\cite{colton} AA is found to have mixed bonding with unique properties. It has got much attention recently as a gas generator due to its weak explosive characteristics.\cite{yakovleva} Using Kamlet-Jacobs equations, the calculated detonation properties, namely, detonation velocity and pressure are found be 6.45 km/s and 15.16 GPa, respectively.\cite{kondal}  AA is used to inflate safety cushions in auto-mobiles with a suitable oxidizer and also used as a solid rocket propellant in photochemical micro-rockets for altitude control.\cite{maycock} In order to address its potential applications, one has to understand the crystal structure and physical properties at the microscopic level. Several experimental studies have been reported in the literature exploring fundamental properties at ambient conditions. Frevel\cite{frevel} first reported the crystal structure of AA which was re-determined later using single crystal\cite{prince} and powder\cite{amorim} X-ray diffraction. The infra-red (IR) vibrational spectra were reported\cite{dows}, which was followed by a combined IR and Raman study.\cite{gray} It was further refined and extended by Iqbal et al\cite{iqbal} who analyzed the vibrational modes of azide ion and the low frequency lattice modes using far-IR and Raman spectroscopic techniques at ambient pressure and low temperatures. From theoretical perspective, the structural, mechanical, electronic and optical properties of AA were reported at ambient pressure from first principles calculations at 0 K.\cite{kondal,Liu1}

\par Since inorganic azides undergo a series of phase transitions involving lattice distortions coupled with changes in the orientation of the linear azide anions thereby forming polymeric nitrogen networks under pressure, high pressure study of these materials has become an active field of research during the last decade.\cite{cheng2} Synthesis of polymeric nitrogen from molecular nitrogen (N$\equiv$N is the strongest bond, 954 kJ/mol) has been achieved\cite{popov,eremets1,hemley} and it consists of individual nitrogen atoms connected by single bonded (N-N) networks. It is the best green high energetic density material (HEDM) known till date.\cite{eremets1, hemley} The azides consist of linear-rod-shaped N$_3^-$ ions with quasi double bonds (N=N bond is weaker than N$\equiv$N, 418 kJ/mol). Hence, it can be expected that the (N=N=N)$^-$ ions of metal azides may form polymeric nitrogen networks more readily than molecular nitrogen (N$\equiv$N).\cite{popov} Among the inorganic azides, AA is quite interesting due to modification in the strength of hydrogen bonding under compression. Raman spectroscopic studies\cite{wu,klapotke1,klapotke2} revealed that it undergoes a polymorphic phase transition around $\sim$3 GPa due to weakening of the hydrogen bonding under pressure. This is reinforced by recent high pressure X-ray and Neutron powder diffraction studies on AA and the transition pressure is found to be $\sim$2.6 GPa,\cite{millar} and $\sim$2.9 GPa,\cite{wu,Wu2,Landerville1} respectively but still the crystal structure of high pressure phase has remained elusive until now. Hu and Zhang\cite{hu} proposed that AA and Trans-Tetrazene (TTZ) could be precursors for the high-pressure synthesis of hydro-nitrogen solid (HNS-1) with $P2_1/m$ crystal symmetry and it can be obtained at 36 and 75 GPa from AA and TTZ, respectively using standard Generalized Gradient Approximation parameterized by Perdew-Burke-Ernzerhof (PBE-GGA) functional. In our previous study\cite{kondal} we systematically studied the effect of semi-empirical dispersion correction methods on structural properties of AA along with standard DFT functionals. The dispersion corrected PBE functional (vdW-TS) developed by Tkatchenko and Scheffler\cite{Tkatchenko1} works better for the molecular crystalline solid AA in contrast to conventional DFT functionals. Following this study, Liu and co-workers predicted the phase stability of N$_4$H$_4$ compounds at ambient\cite{Liu2} as well as under compression up to 80 GPa\cite{Liu3} using the vdW-TS method. They proposed a new tetragonal phase (HNS-2) with crystal symmetry $P4_2/n$. The order of stability of the four phases at ambient pressure is as follows: AA $>$ TTZ $>$ HNS-2 $>$ HNS-1 and they also encounter a series of phase transitions from AA $\rightarrow$ HNS-2 at 5.6 GPa, HNS-2 $\rightarrow$ AA at 15 GPa, AA $\rightarrow$ TTZ at 30 GPa and TTZ $\rightarrow$ HNS-1 at 69.2 GPa. Recently these transitions are revisited for AA, TTZ, HNS-1 phases and reported\cite{Landerville1} that the transitions found to occur between AA $\rightarrow$ TTZ at 41.4 GPa and TTZ $\rightarrow$ HNS-1 at 89.4 GPa, respectively using DFT-D2 method. There is an inconsistency between theoretical calculations in predicting the phase transition pressures using various dispersion corrected approaches and also dependence of temperature on phase stability of N$_4$H$_4$ compounds is unknown. With this motivation, we have investigated relative phase stability of N$_4$H$_4$ compounds, structure and lattice dynamics of AA at ambient as well as at high pressure using dispersion corrected density functional calculations. The rest of the paper is organized as follows: in section II, we briefly describe the methodology of computation. Results and discussion concerning phase stability of hydro-nitrogen solids as well as structural and lattice dynamics of AA under high pressure are presented in section III. Finally section IV summarizes the conclusions of the paper.

\section{Method of computation}
Thermodynamic stability was investigated by comparing free energy of the investigated hydro-nitrogen solids. In order to predict the thermodynamic ground state of N$_4$H$_4$ compounds, we have calculated Gibbs free energy of the four investigated compounds. The Gibbs free energy for any material at a given temperature is
\begin{equation}
G(P, T) = F(V, T) + PV = F_{vib} + F_{perfect} + PV,
\end{equation}
where F, P, T and V are the Helmholtz free energy, pressure, absolute temperature and volume, respectively. F(V,T) is the sum of vibrational free energy and perfect lattice energy \emph{i.e.} F(V, T) = F$_{vib}$ + F$_{perfect}$. The vibrational free energy F$_{vib}$ calculated within the harmonic approximation is given by
\begin{equation}
 F_{vib} = k_BT\sum\limits_{i}\Big{(}\frac{\hslash\omega_i}{2k_BT} + ln(1-e^{\frac{-\hslash\omega_i}{k_BT}})\Big{)},
\end{equation}
where $\omega_i$ is the $i^{th}$ phonon frequency, $\hslash$ is the reduced Planck constant and $k_B$ is the Boltzmann constant, and F$_{perfect}$ = E$_0$ + E$_{el}$ - TS$_{el}$. E$_0$ and E$_{el}$- TS$_{el}$ are the contributions from the lattice and electronic excitations, respectively. If the electronic excitations are neglected then the Gibbs free energy can be calculated by G =  E$_{0}$ + PV + F$_{vib}$ = H$_{0}$ + F$_{vib}$, where H$_{0}$ = E$_{0}$ + PV is the enthalpy at T = 0 K and E$_0$ denotes the change in energy of atoms in the unit cell of a perfect crystal with respect to the energy of the isolated atoms.

The van der Waals (vdW) interactions missing in the standard DFT can be efficiently modeled by adding a pair wise interatomic interaction (C$_6R^{-6}$) to the DFT energy \\
\begin{equation}
E_{vdW} = -\frac{1}{2}\sum\limits_{A, B}\frac{C_{6AB}}{R_{AB}^6}f_{damp}(R_{AB}, R_A^0, R_B^0),
\end{equation}
where R$_{AB}$ is the distance between atoms A and B, C$_{6AB}$ is the corresponding C$_6$ coefficient, $R_A^0$ and $R_B^0$ are the vdW radii of atoms A and B. \\
One of the novel methods for computing non-empirical dispersion coefficients is vdW-TS scheme of Tkatchenko and Scheffler,\cite{Tkatchenko1} which is sensitive to chemical environment of the atoms. In this method, C$_{6AB}$ is obtained from the Casimir-Polder integral\cite{} which is given by \\
\begin{equation}
C_{6AB} = \frac{2C_{6AA}C_{6BB}}{\frac{\alpha_0^B}{\alpha_0^A}C_{6AA}+\frac{\alpha_0^A}{\alpha_0^B}C_{6BB}}
\end{equation}
and a Fermi type damping function f$_{damp}$ = $\frac{1}{1+e^{-d(R_{AB}/R_0-1)}}$ is used to eliminate spurious interactions at too short distances. The free-atom reference values of $\alpha_0^A$ and $C_{6AA}$ are taken from the self-interaction corrected Time Dependent(TD)-DFT calculations of Chu and Dalgarno.\cite{Chu} They take advantage of the relationship between the effective volume and polarizability to calculate dispersion coefficients that depend on the chemical environment of the atom. Hirshfeld partitioning\cite{Hirshfeld} of the electron density of the system is used to obtain contribution from each atom to the density. This effective density, and hence the volume, is compared to the density of the free-reference atom to obtain a scaling factor which is used to define the response of the dispersion coefficients to chemical environment. However, the vdW-TS scheme does not include long range electrostatic screening beyond the range of the exponentially decaying atomic densities.\cite{Tkatchenko2} Recently Tkatchenko et al\cite{Tkatchenko2} proposed a computationally efficient method (TS-SCS) extension to the vdW-TS method, in which they considered electrodynamic response effects, in particular, the interactions of the atoms with the dynamic electric field due to the surrounding polarizable atoms and also account for many-body effects based on the coupled fluctuating dipole model. In this method, the frequency dependent screened polarizability is obtained by solving self-consistent screening equation.
\begin{equation}
\alpha_A^{SCS}(i\omega) = \alpha_A^{TS}(i\omega) - \alpha_A^{TS}(i\omega)\sum\limits_{B\neq A} \tau_{A,B}\alpha_B^{SCS}(i\omega).
\end{equation}
\par Cambridge Series of Total Energy Package (CASTEP)\cite{Payne} is used to calculate the structural and vibrational spectrum of the N$_4$H$_4$ compounds. We have used Norm-conserving (NC) pseudo potentials (PP) for electron-ion interactions, as they are well suited for phonon calculations\cite{Troullier} while PBE-GGA\cite{Burke} was used to treat electron-electron interactions. The Broyden-Fletcher-Goldfarb-Shanno (BFGS) minimization scheme\cite{Almlof} has been used for the structural relaxation. The convergence criteria for structural optimization was set to ultrafine quality with a kinetic energy cutoff of 850 eV and 2$\pi\times$0.025$\textup{\AA}^{-1}$ separation of k-mesh according to the Monkhorst-Pack grid scheme.\cite{Monkhorst} The self-consistent energy convergence less than 5.0$\times$10$^{-6}$ eV/atom and maximal force between atoms was set to 0.01 eV/$\textup{\AA}$. The maximum displacement and stress were set to be 5.0$\times$10$^{-4}\textup{\AA}$ and 0.02 GPa, respectively.

The elastic constants and phonon dispersion curves of AA are computed using density functional perturbation (DFPT)\cite{Baroni} theory implemented in the Vienna ab-initio Simulation Package (VASP)\cite{Kresse}. The initial geometry is further relaxed, and ensured a force convergence of the order 10$^{-5}$ eV/atom with zero external pressure. The phonon dispersion that is calculated using the unit cell shows the frequencies of some of the modes to be imaginary. Since VASP code computes the dynamical matrix in real space, we have to use a sufficiently large super cell to incorporate the long range nature of the dynamical matrix.  To compute the phonon dispersion curves a super cell of size 1 $\times$ 2 $\times$ 1 (64 atoms) has been used. The force constant matrix is computed and the dynamical matrix is obtained from its Fourier transform. The eigenvalues and eigenvectors of the dynamical matrix will yield square of the phonon frequencies and polarization of each mode. The DFPT calculations will give the phonon frequencies and polarization at zone center (q=0). In order to obtain the phonon dispersion in the entire Brillouin zone, we used an auxiliary post process package PHONOPY.\cite{Togo} All the CASTEP and VASP calculations were performed using vdW-TS and TS-SCS schemes, respectively.
\section{Results and discussion}
\subsection{Phase stability of N$_4$H$_4$ compounds}
\begin{figure}[h]
\centering
\includegraphics[height = 3.7in, width=5.5in]{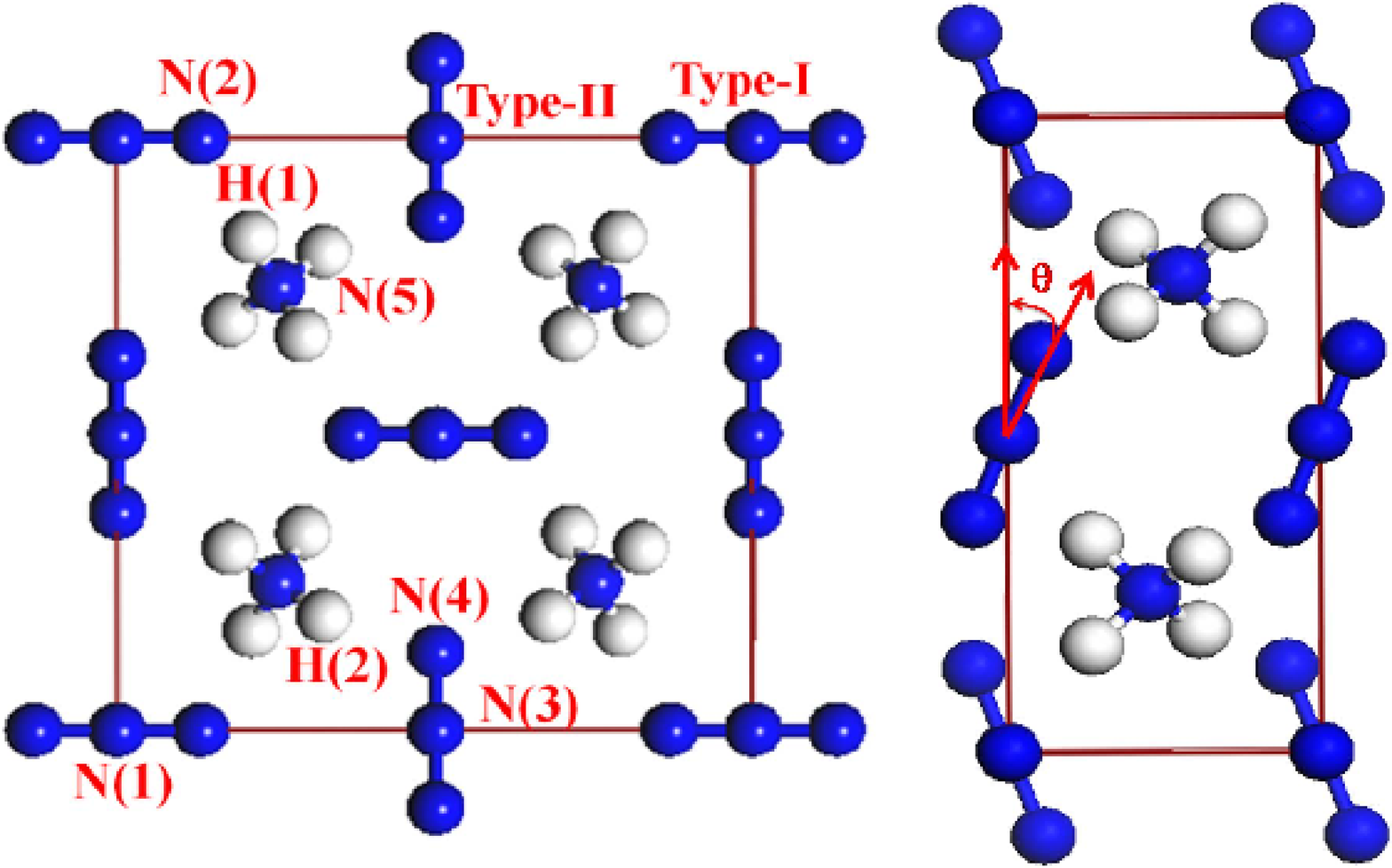}
\caption{(Colour online) Crystal structure of AA viewed in a-c (left) b-c (right) planes. Azide (N$_3$) ion parallel and perpendicular to a-axis named as Type I and Type-II, respectively. Angle $\theta$ is defined as orientation of Type-II azides w.r.t. to c-axis.}
\label{fig:AA}
\end{figure}

\begin{figure}[h]
\centering
\includegraphics[height = 6.0in, width=6.0in]{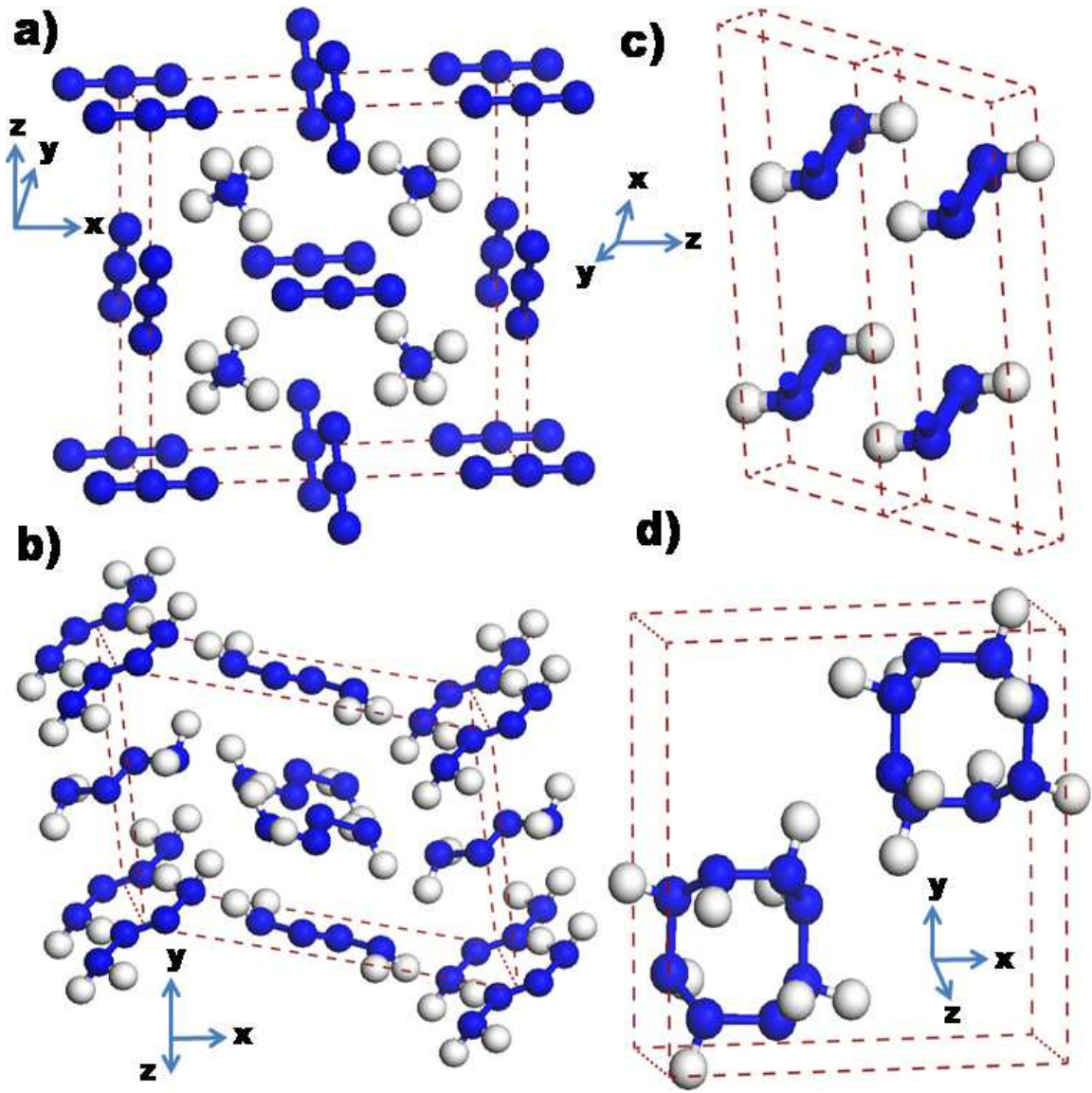}
\caption{(Colour online) Crystal structures of hydro-nitrogen solids; (a) AA, (b) TTZ, (c) HNS-1 (super cell) and (d) HNS-2.}
\label{fig:NH}
\end{figure}

At ambient conditions, AA crystallizes in the orthorhombic structure with space group $Pmna$.\cite{wu,prince,amorim,millar} It possesses distorted CsCl-type structure because of strong hydrogen bonding between NH$_4^+$ and N$_3^-$ ions and the arrangement of azide ion is similar to that in the ambient structures of heavy metal (Ag, Tl) azides. Due to presence of the hydrogen bonding one half of the azide anions lie parallel to a-axis (Type-I) and the other half lie perpendicular to a-axis (Type-II)  as depicted in figure \ref{fig:AA} by allowing a tetrahedral arrangement around a single ammonium cation.\cite{millar} TTZ is a molecular crystal, which consists of non-linear arrangement of four N-atoms in a chain with two hydrogen atoms each attached to both ends of the N$_4$ molecule and four such molecules are arranged with $P\bar{1}$ symmetry in the unit cell.\cite{Veith} HNS-1 contains similar molecular structure as TTZ with a hydrogen atom attached to each nitrogen atom in each (NH)$_2$ unit and the two units are separated by a distance within the unit cell of $P2_1/m$ symmetry.\cite{hu} Recently a new tetragonal phase has been predicted with $P4_2/n$ symmetry,\cite{Liu3} where two (NH)$_4$ units form a distorted cyclic octagon within the unit cell, in which each nitrogen atom is attached to one hydrogen atom. The crystal structures of the four N$_4$H$_4$ compounds are shown in figure \ref{fig:NH}  and all these phases are molecular crystalline solids. However, most of the energetic systems such as primary (Fulminates, Azides, Styphanates, etc.) and secondary (RDX, HMX, TNT, PETN, etc.) explosive molecular crystals are bonded through weak dispersive interactions. The non-covalent weak interactions such as hydrogen bonding and dispersive interactions play a significant role in determining the stability of these molecular solids. It is well known that the standard DFT functionals are inadequate to predict the long range interactions in molecular crystalline solids. Consequently, semi-empirical dispersion correction methods have been developed and implemented through the standard DFT functionals to provide an accurate description of long range weak interactions. In our previous study\cite{kondal}, we have investigated the structural and elastic properties of AA at ambient pressure using the dispersion corrected DFT-D2 and vdW-TS methods along with standard DFT functionals. As discussed in the previous section, we also show that the vdW-TS method is successful in predicting the ground state properties of AA over DFT-D2 method.\cite{kondal, Landerville2} Hence  we have used the dispersion corrected vdW-TS and TS-SCS methods for all the calculations in the present study. As a first step, we have performed full structural optimization including lattice geometry and fractional coordinates of the four N$_4$H$_4$ compounds. The calculated unit cell parameters for these compounds are consistent with the available experimental data\cite{wu,prince,amorim,millar} and other theoretical calculations\cite{hu,Liu2,Liu3} which are presented in Table \ref{tab:lp}.
\par
\begin{figure}[h]
\centering
\includegraphics[height = 4.0in, width=5.5in]{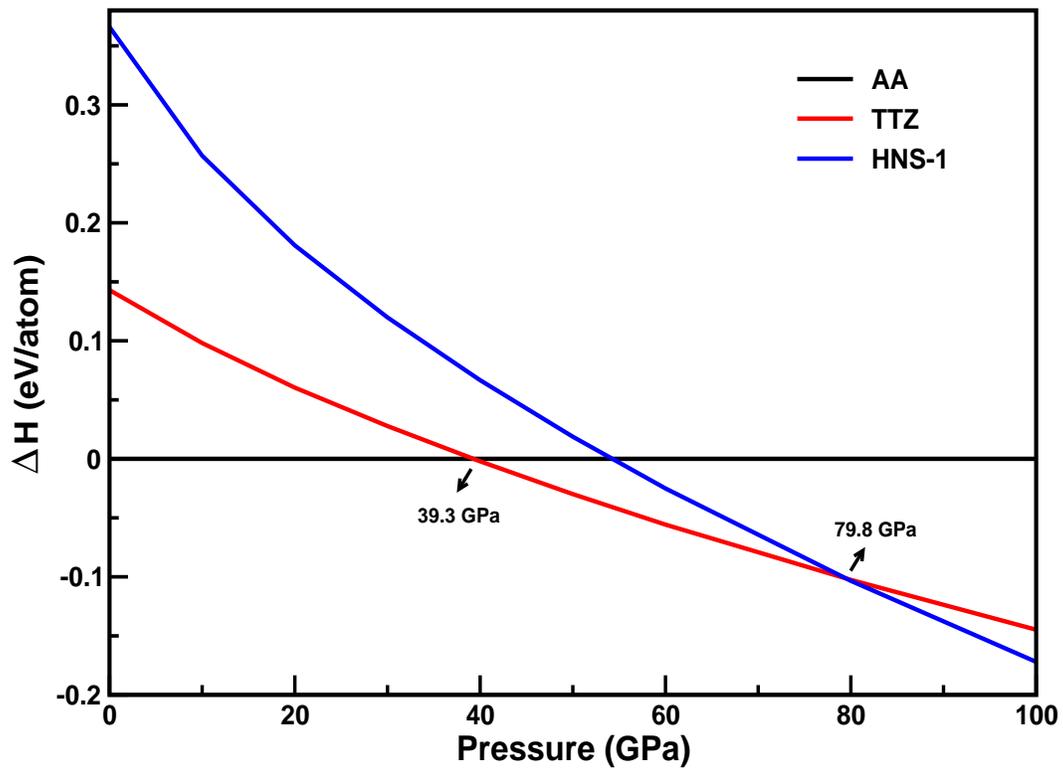}
\caption{(Colour online) Calculated relative enthalpies of TTZ and HNS-1 with respect to AA using NCPP approach as a function of pressure.}
\label{fig:PT_P}
\end{figure}

 From our first principles calculations, we found that total energies of the N$_4$H$_4$ compounds increase in the following order; E$_0^{AA}$ $<$ E$_0^{TTZ}$ $<$ E$_0^{HNS-2}$ $<$ E$_0^{HNS-1}$ at ambient pressure. The total energy of the AA is lowered by 0.15, 0.36, 0.38 eV/atom when compared to TTZ, HNS-2, and HNS-1, respectively and the order of stability is consistent with the previous calculations.\cite{hu,Liu2,Liu3} This clearly shows that AA is the relatively stable phase among the studied N$_4$H$_4$ compounds at ambient pressure. What happens to the order of stability at higher pressures for these four phases$?$ We made an attempt to investigate the phase transitions and the possibility for the formation of hydro-nitro solid under high pressure. As illustrated in figure \ref{fig:PT_P}, we predict a series of phase transitions AA $\rightarrow$ TTZ at 39.3 GPa, TTZ $\rightarrow$ HNS-1 at 79.8 GPa using NCPP approach, which are in good agreement with recent theoretical calculations\cite{Landerville1} using dispersion corrected DFT-D2 method. Also the transition pressures obtained using ultra-soft pseudo-potential (USPP) approach are found to be AA $\rightarrow$ TTZ at 27.5 GPa, TTZ $\rightarrow$ HNS-1 at 74.2 GPa (see figure 1 of the supplementary material) and are comparable with the ones obtained in the study by Liu et al.\cite{Liu3} The inconsistency between transition pressures of the examined compounds are due to difference in the pseudo-potentials used in the calculations. Recent Raman spectroscopic measurements reveal that the HNS-1 will be formed at even much higher pressures ($>$ 71 GPa) and hence the transition pressures obtained using NCPP including dispersive interactions are in good agreement with the experimental observations.\cite{Landerville1} Medvedev et al\cite{klapotke1} determined the pressure at which polymorphic structural phase transition from phase I to II occurs at around 3 GPa and phase II consists of NH$_4$ and N$_3$ ions similar to phase I with a difference that both of the azide groups must occupy equivalent crystallographic positions to be consistent with observations from Raman spectroscopy. This is further confirmed by the single crystal X-ray diffraction study and the phase II is temporarily assigned as tetragonal phase.\cite{wu} Recently Liu et al\cite{Liu3} predicted the crystal structure of the tetragonal phase with $P4_2/n$ symmetry and it possesses a distorted cyclic structure which is in contrast to the structure proposed by the Raman spectroscopic measurements.\cite{klapotke1} In addition, they also find structural transitions from AA $\rightarrow$ HNS-2 at 5.6 GPa and HNS-2 $\rightarrow$ AA at 15 GPa. In order to resolve the above issues we systematically investigated the high pressure behavior of AA and HNS-2 phases and we found that AA does not undergo a transition to HNS-2 phase below 20 GPa (see figure 2 of the supplementary material). Therefore the prediction/determination of crystal structure of phase II is still an open challenge for both theoreticians and experimentalists.
\par
\begin{figure}[h]
\centering
{\subfigure[]{\includegraphics[height = 2.0in, width=3.5in]{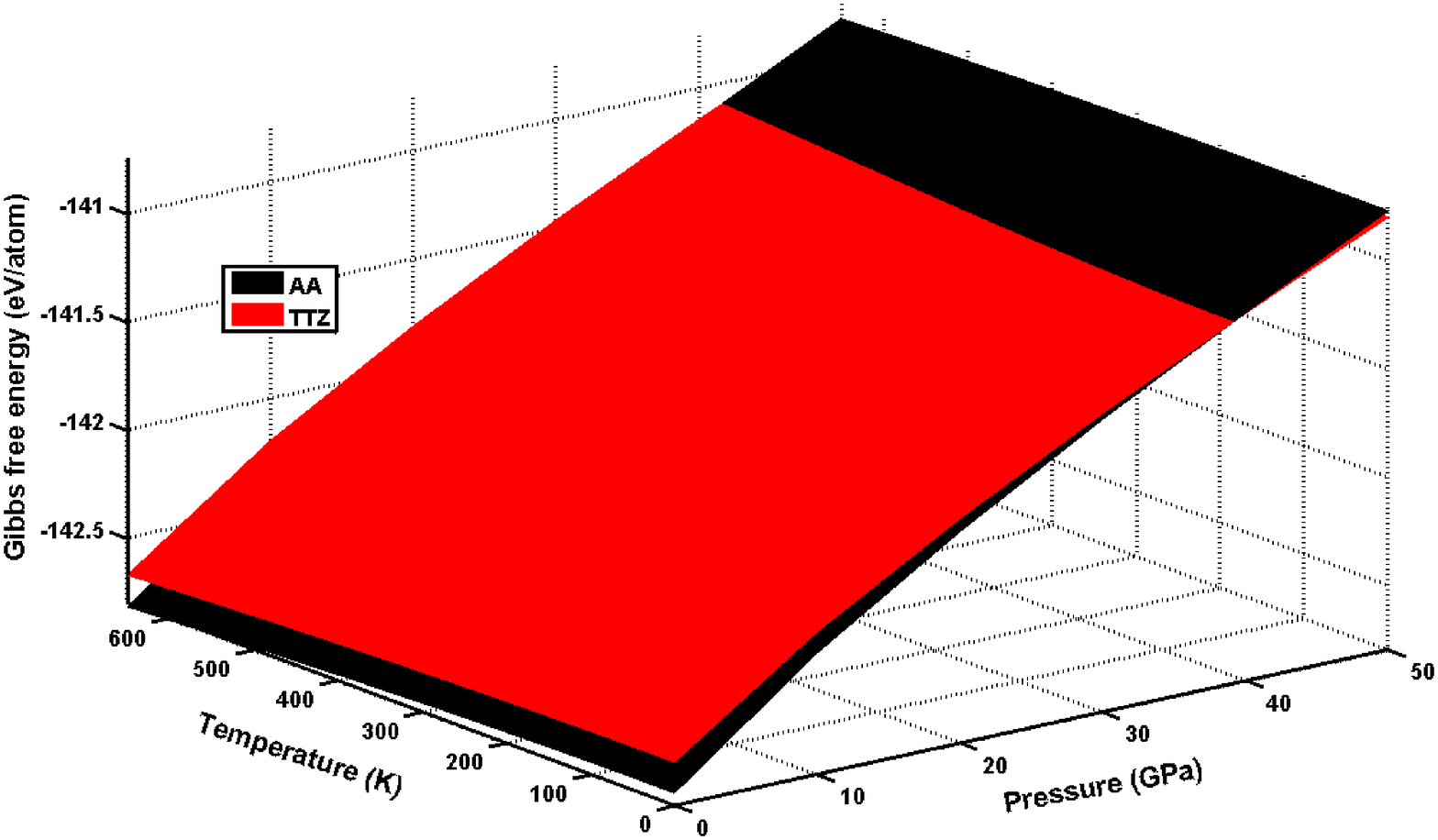}}}
{\subfigure[]{\includegraphics[height = 1.8in, width=2.8in]{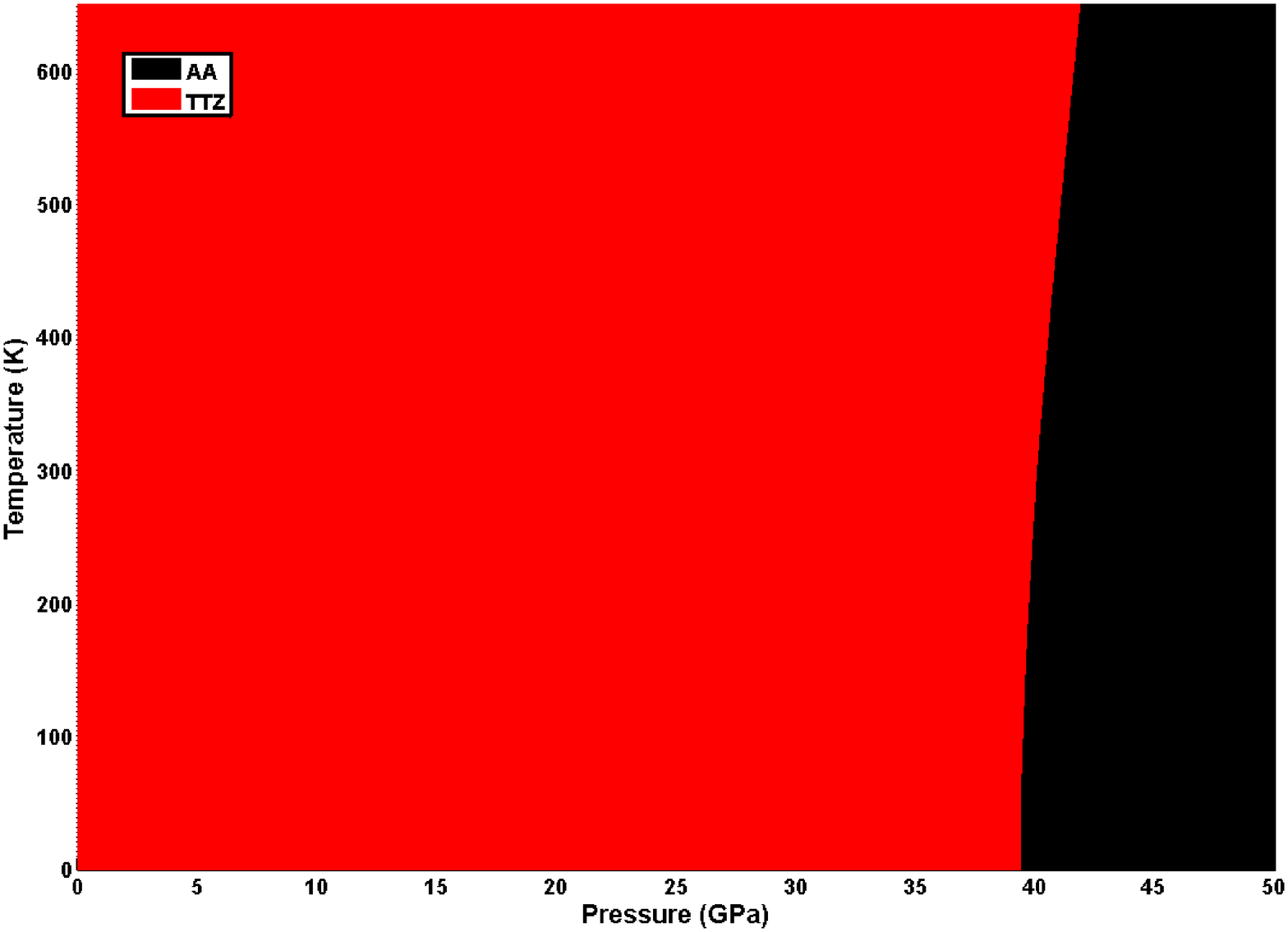}}}
{\subfigure[]{\includegraphics[height = 2.0in, width=3.5in]{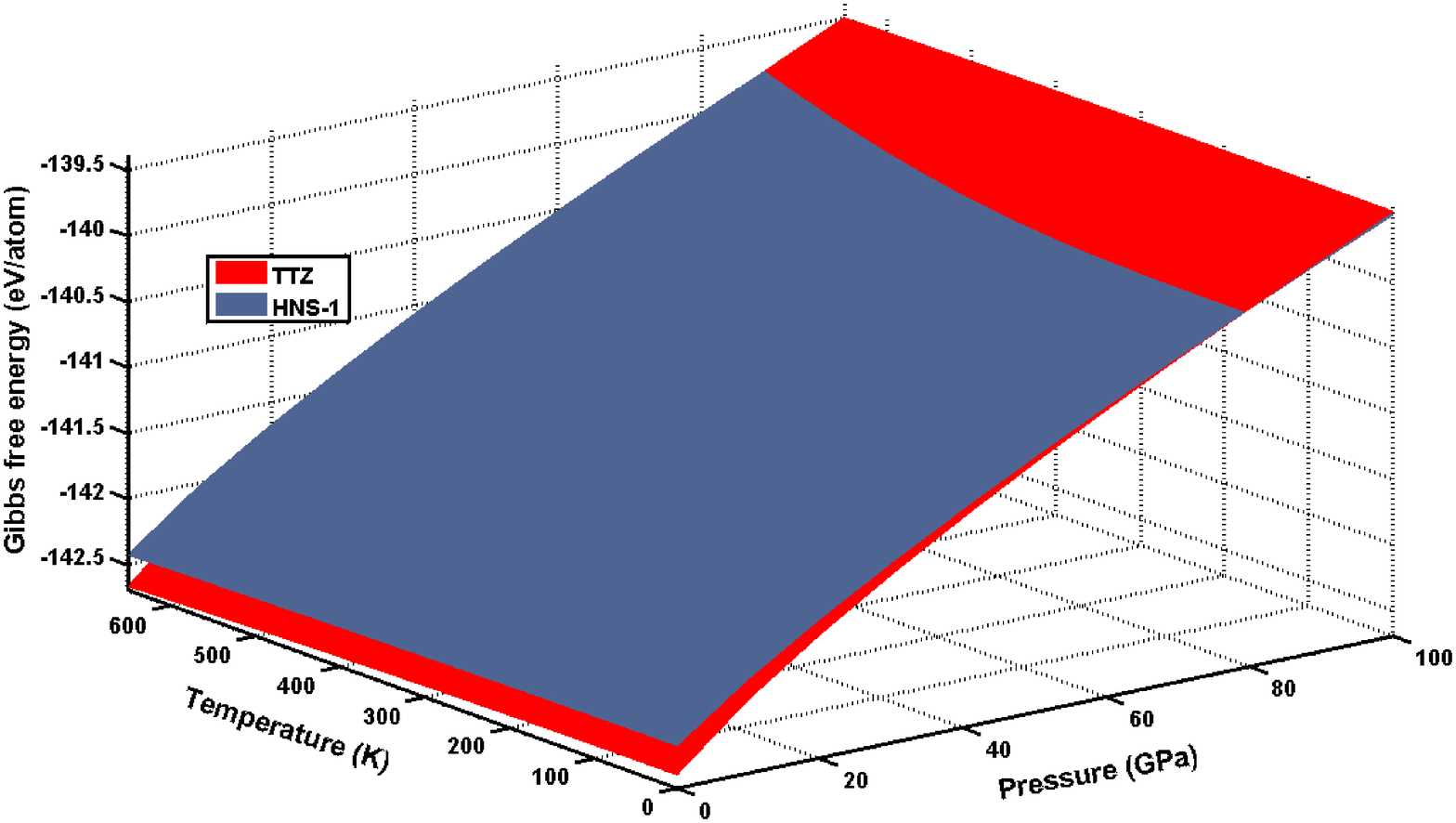}}}
{\subfigure[]{\includegraphics[height = 1.8in, width=2.8in]{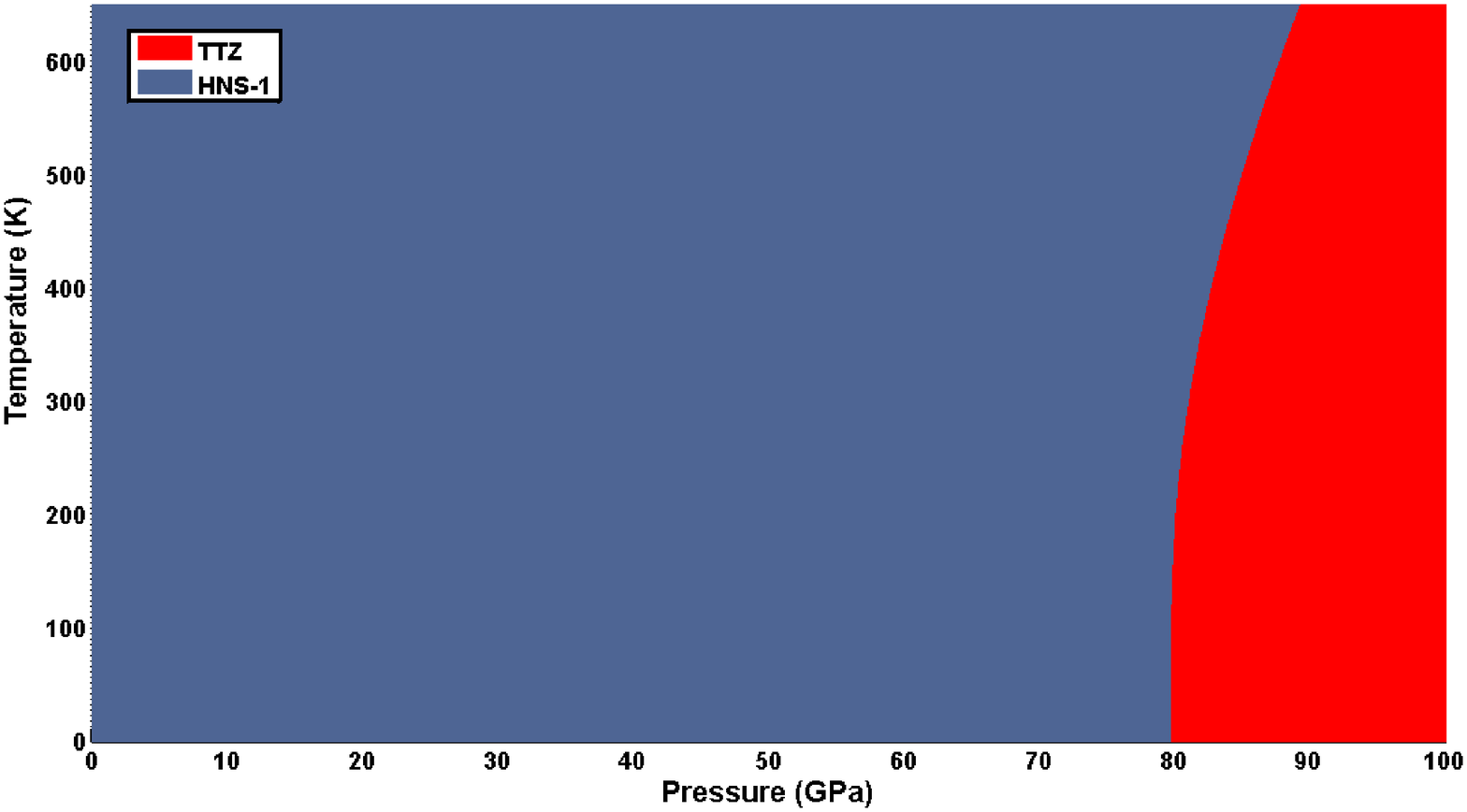}}}
\caption{(Colour online) Calculated phase diagram (G, P, T) of (a) AA and TTZ, (b) viewed in P-T plane for AA and TTZ (c) TTZ and HNS-1, and (d) viewed in P-T plane for TTZ and HNS-1 using NCPP approach.}
\label{fig:PT_PT}
\end{figure}

 In addition, we also considered the contribution of lattice vibration at elevated pressure and temperatures. Since AA decomposes at 673 K (400 $^0$C)\cite{yakovleva}, we have plotted Gibbs free energy as a function of pressure (0-100 GPa) and temperature (0-650 K) within vdW-TS method. As illustrated in figure \ref{fig:PT_PT}, AA undergoes a structural transition to TTZ around 39.3 GPa at 0 K (see figure \ref{fig:PT_P}) and the transition pressure increases to 42.6 GPa with increasing temperature up to 650 K. Further, TTZ transforms to HNS-1 around 79.8 GPa at 0 K (see figure \ref{fig:PT_P}) and the transition pressure increases from 79.8 to 89.7 GPa when the temperature raises from 0 to 650 K. Overall, we observe that AA is the thermodynamic ground state of N$_4$H$_4$ {\bf compounds} which is consistent with the previous theoretical calculations\cite{Liu2,Liu3,Wu2} and also AA transforms to TTZ around $\sim$ 39-43 GPa followed by transformation of TTZ to HNS-1 around $\sim$ 80-90 GPa under the studied temperature range 0-650 K.

\subsection{Structural and mechanical properties of AA under high pressure}
We have also studied the structural properties of the thermodynamic ground state (AA) of N$_4$H$_4$ compounds under pressure up to 0-2.5 GPa, 0-20 GPa  and 20-50 GPa in steps of 0.25, 2 and 5 GPa, respectively. The pressure dependent unit cell parameters obtained using vdW-TS method are in good accordance with the experimental data\cite{wu,millar} within 0-2.5 GPa with an exception for lattice parameter c, which shows a larger contraction due to over binding within the studied pressure range (see figure \ref{fig:lattice}a). As shown in figure \ref{fig:lattice}b, we observe that there is a strong anisotropy between the lattice constants $a$, $b$ and $c$ in the low pressure region (0-3 GPa). The lattice constants $a$ and $c$ approach towards each other because of accelerated and decelerated compression of $a$ and $c$ lattice constants respectively and they merge between 18-22 GPa. This is due to rotation of Type-II azide ions and reduction in the hydrogen bond energy,\cite{Wu2} consequently the anisotropic phase I undergoes transition to phase II under compression. This will be further confirmed by lattice dynamical calculations under pressure in the up coming section. Raman spectroscopic studies suggest that phase II is stable up to 55 GPa\cite{klapotke1}. However, based on our present calculations we predict that phase II further transforms to a less anisotropic low symmetry structure around 22 GPa. In addition, we have also compared the calculated normalized lattice parameters ($a$, $b$, $c$, and $V$) at ambient pressure (0 GPa) with the experimental data (0.07 GPa is experimental ambient pressure).\cite{millar} The calculated relative compressibilities of unit cell axes are presented in figure \ref{fig:lattice}c along with the experimental results.\cite{millar} The reduction of $b$ lattice parameter is larger under compression which is clearly seen from the study of the first order pressure coefficients $\gamma($X$)$ = $\frac{1}{X}\frac{dX}{dP}$ ($X$ = $a$, $b$, $c$). The calculated first order pressure coefficients (in GPa$^{-1}$) using vdW-TS scheme are found to be 8.9 $\times$ 10$^{-3}$, 20.4 $\times$ 10$^{-3}$, and 6.6 $\times$ 10$^{-3}$ for lattice parameters $a$, $b$, and $c$, respectively. This clearly shows that $b$-axis is the most compressible due to flattening of the wave-like planes of N$_3^-$ anions under compression which further reveals that AA is sensitive to impact along the $b$-axis. Crystallographic $c$-axis is the least compressible because of Type-II azide anions align parallel to $c$-axis with increasing pressure as shown in figure \ref{fig:lattice}c, resulting in an excellent agreement with the experimental observation\cite{wu,millar} as well as with our recent theoretical prediction from the longitudinal elastic constants.\cite{kondal} The calculated ordering of the elastic constants (C$_{33}$ $\textgreater$ C$_{11}$ $\textgreater$ C$_{22}$) further confirm that $b$-axis is the most and $c$-axis is the least compressible axes for AA.\cite{kondal,Liu2} As illustrated in figure \ref{fig:lattice}d, the calculated pressure-volume (P-V) data is consistent with the experiments\cite{millar,laura} especially in the low pressure region. This might be due to fact that the dispersion coefficients C$_6$ used for the chemical species N and H are kept constant over the studied pressure range. Overall, the experimental trends are well reproduced by the vdW-TS method up to 2.5 GPa. We also calculated equilibrium bulk modulus ($B_0$) and it is found to be 27.6 GPa by fitting pressure-volume data to 2$^{nd}$ order Birch-Murnaghan equation of state.\cite{murnaghan} The obtained $B_0$ value of 27.6 GPa is slightly higher when compared to the experimental values of 24.5\cite{wu}, 20.2\cite{millar}, and 19.8\cite{laura} GPa and it is consistent with our previously calculated value of 26.34 GPa using USPP.\cite{kondal} The $B_0$ of AA lies between the $B_0$ values of Alkali Metal Azides (AMA) and Heavy Metal Azides (HMA) indicating that AA is relatively harder than AMA and softer than HMA.

The elastic behavior of a crystalline solid is described by its matrix of second-order elastic constants, which is given in the Voight notation by the expression
\begin{equation}
C_{ij} = \frac{1}{V}\left(\frac{\partial^2E}{\partial\epsilon_i\partial\epsilon_j}\right),
\end{equation}
where V is the equilibrium volume, $\epsilon$ denotes the strain and E is total energy of the crystal which is given by
\begin{equation}
E = E_0 + \frac{1}{2}V \sum\limits_{i,j=1}^6 C_{ij}\epsilon_i\epsilon_j + O(\epsilon^3),
\end{equation}
where E$_0$ is the energy of the unstrained perfect crystal. The elastic constants matrix is a real symmetric matrix of size 6$\times$6,\cite{Mouhat} and hence it can have at most 21 independent matrix elements. These 21 matrix elements are indeed independent of each other in a crystal with least symmetry. Thus there are 21 independent elastic constants for crystals with triclinic Bravais lattice. The number of independent elastic constants become smaller as the symmetry of the crystal increases. There are only three independent elastic constants for crystals with cubic symmetry. It turns out that there are 9 independent elastic constants for crystals with orthorhombic structure, which are C$_{11}$, C$_{22}$, C$_{33}$, C$_{44}$, C$_{55}$, C$_{66}$, C$_{12}$, C$_{13}$, and C$_{23}$. Further, the nine elastic constants are classified into three categories such as (i) longitudinal ( C$_{11}$, C$_{22}$, C$_{33}$), (ii) shear (C$_{44}$, C$_{55}$, C$_{66}$) and (iii) transverse coupling (C$_{12}$, C$_{13}$, C$_{23}$) elastic constants. C$_{11}$, C$_{22}$, and C$_{33}$ couple an applied normal stress component in the crystallographic $a$, $b$ and $c$-directions with a uniaxial strain along the crystallographic $a$, $b$ and $c$-axes, respectively. On the other hand, C$_{44}$, C$_{55}$, and C$_{66}$ couple an applied shear stress in the crystallographic $bc$, $ac$ and $ab$-planes with a shear strain along the crystallographic $bc$, $ac$ and $ab$-planes, respectively. Finally, C$_{12}$ and C$_{13}$ couple an applied normal stress component in the crystallographic $a$-direction with a uniaxial strain along the crystallographic $b$ and $c$-axes, respectively whereas C$_{23}$ couples a uniaxial strain along $c$-direction to an applied normal stress component along crystallographic $b$-direction.

In order to get more insight on structural phase transition in AA, we have also calculated the elastic constants at ambient as well as at high pressure up to 20 GPa. The calculated elastic constants at ambient pressure are consistent with our previous\cite{kondal} and other\cite{Liu2} theoretical calculations. A crystalline lattice is mechanically stable at zero pressure only if the Born stability criteria\cite{Born} are fulfilled. In case of orthorhombic systems the stability criteria are\cite{Mouhat}
\begin{gather}
C_{11} > 0, C_{11}C_{22} > C_{12}^2,  \nonumber \\
C_{11}C_{22}C_{33}+2C_{12}C_{13}C_{23}-C_{11}C_{23}^2-C_{22}C_{13}^2-C_{33}C_{12}^2 > 0, \nonumber \\
C_{44} > 0, C_{55} > 0, C_{66} > 0
\end{gather}
The obtained elastic constants satisfy the Born stability criteria indicating that AA is mechanically stable at ambient pressure. When a non-zero uniform stress is applied to the crystal, the above criteria to describe the stability limits of the crystal at finite strain are not adequate and the Born stability criteria must be modified.\cite{Monteseguro} Sin'ko and Smirnov reported the theoretical conditions of elasticity under pressure\cite{Sinko1,Sinko2} and the modified elastic constants for a orthorhombic crystal under pressure are \~{C$_{ii}$} = C$_{ii}$-P, (for $i$ = 1-6), \~{C$_{12}$} = C$_{12}$+P, \~{C$_{13}$} = C$_{13}$+P and \~{C$_{23}$} = C$_{23}$+P.
Hence new stability criteria are obtained by replacing the elastic constants in eqn (8) by the modified elastic constants. Therefore, AA is mechanically stable under hydrostatic pressure when the generalized Born stability criteria are:
\begin{gather}
M1 = C_{11}-P > 0, M2 = C_{11}C_{22}-P(C_{11}+C_{22}) > 2PC_{12}+C_{12}^2,  \nonumber \\
M3 = C_{11}C_{22}C_{33}+2C_{12}C_{13}C_{23}-C_{11}C_{23}^2-C_{22}C_{13}^2-C_{33}C_{12}^2 \nonumber  \\
      +P(C_{12}^2+C_{13}^2+C_{23}^2-C_{11}C_{22}-C_{22}C_{33}-C_{33}C_{11})  \nonumber \\
      +2P(C_{12}C_{13}+C_{12}C_{23}+C_{13}C_{23}-C_{11}C_{23}-C_{22}C_{13}-C_{33}C_{12}) \nonumber  \\
      +4P^2(C_{12}+C_{13}+C_{23})+4P^3 > 0, \nonumber \\
M4 = C_{44}-P > 0, M5 = C_{55}-P > 0, M6 = C_{66}-P > 0
\end{gather}
are simultaneously satisfied. As depicted in figure \ref{fig:elastic}a, the calculated elastic constants show a non-monotonic variation with pressure. The pressure
dependent elastic constants obey the generalized Born stability criteria as given in eqn (9) except M5 and it is plotted as a function of pressure along with M6. As illustrated in figure \ref{fig:elastic}b, M6 decreases up to 5 GPa and later it increases up to 16 GPa and then starts decreasing with pressure, which indicates that the applied pressure drives shear instability in the system up to 5 GPa and then stabilizing above 5 GPa. Also M5 violates the stability criteria between 18-20 GPa, which clearly indicates the mechanical instability of AA under high pressure. Softening of the shear elastic moduli M5 and M6 suggests a shear instability of AA under pressure. Karki et al\cite{Karki} reported that elastic instability bounds the transition pressure and determines the transition pressure precisely for first and second order phase transitions, respectively. Mechanical stability criteria, M6 and M5 bounds the transition pressure for AA and hence one can expect that the phase transition in AA is of first order type.
\subsection{Hydrogen bonding in AA under pressure}
As discussed in section I, Raman spectroscopic measurements on AA\cite{klapotke1,klapotke2} revealed that it undergoes a polymorphic structural phase transition about 3.0 GPa due to weakening of the hydrogen bonding under pressure. They also proposed two possible reasons for weakening of hydrogen bonding\cite{klapotke1} which are 1) elongation of N...N distance due to the rotation of azide (N$_3^-$) ion, which leads to increment of the N-H bond length. 2) Reorientation of NH$_4^+$ ions leading to bending of N-H...N bond.\cite{klapotke1} A recent experimental study \cite{wu} on AA shows that N-H...N bond decreases with pressure and the weakening of hydrogen bonding is only because of the bending of N-H...N bond, which is possibly due to movement of ammonia cation and rotation of azide (Type-II) ions. In order to confirm this, we made an attempt to investigate the high pressure behavior of intermolecular hydrogen bonds. The angle made by Type-II azides with c-axis (see figure \ref{fig:AA}), ${i.e.}$ angle $\theta$, is plotted in the pressure range 0-2.5 GPa. As seen from figure \ref{fig:beta}a, the angle $\theta$ decreases with pressure, which is consistent with the experimental data\cite{wu} whereas the Neutron powder diffraction data shows a much rapid decrease of the angle in the pressure range of 0-2.5 GPa.\cite{millar} Further we investigated the angle $\theta$ beyond 2.5 GPa and observe a rapid decrease below 12 GPa, it reaches a minimum value at 12 GPa and then it starts increasing with pressure up to 50 GPa as shown in figure \ref{fig:beta}b which is due to re-orientation of Type-II azide ions with respect to c-axis under compression. The calculated hydrogen bond lengths (N(5)...N(2) and N(5)...N(4)) and the corresponding bond angles (N(5)-H(1)...N(2) and N(5)-H(2)...N(4)) under pressure are shown in figure 4 of the supplementary material. The calculated N(5)...N(2) and N(5)...N(4) bond lengths decrease with pressure and are in good agreement with the experimental data up to 2.5 GPa.\cite{wu} The reorientation of NH$_4^+$ ions leads to bending of N-H...N hydrogen bond and this can be analyzed by calculating the intermolecular hydrogen bond angles of N-H...N bonds under compression. The calculated hydrogen bond angles show a significant variation ($\sim$176-178) between 4-10 GPa and the N(5)-H(2)...N(4) reaches a maximum value at 10 GPa due to orientation of Type-II azide ions as shown in figure \ref{fig:BA_H}.

\subsection{Zone centre vibrational spectrum of AA at ambient pressure}
IR and Raman spectroscopic techniques have been used for the identification and characterization of energetic materials.\cite{fell2} The vibrational properties of a molecular solid provide information about the intra- and inter-molecular bonding. Therefore, the changes in vibrational properties are important indicators for understanding physical and chemical changes in the molecular crystals. The quantum mechanical DFT methods have proven themselves as essential tools for interpretations and predicting vibrational spectra of materials.\cite{hess,krishnakumar} In the present study, the vibrational spectrum of AA is calculated using linear response method within density functional perturbation theory (DFPT) and a detailed analysis of the vibrational spectra and their complete assignment have been made at ambient as well as at high pressure. Single crystal X-ray and Neutron powder diffraction studies\cite{wu,prince,amorim,millar} revealed that AA crystallizes in the orthorhombic structure with \emph{Pmna} space group; NH$_4^+$ and N$_3^-$ ions occupy the sites of symmetry C$_2$ and C$_{2h}$, respectively. It consists of four molecules per unit cell (or 32 atoms per cell) resulting in the 96 vibrational modes, which are further classified into 3 acoustic and 93 optical modes. According to the group theory analysis of \emph{Pmna} space group, the symmetry decomposition of the modes as follows. \\
$\Gamma_{acoustic}$ = B$_{1u}$ $\oplus$ B$_{2u}$ $\oplus$ B$_{3u}$   \\
$\Gamma_{optical}$  = 15B$_{1u}$ $\oplus$ 14B$_{2u}$ $\oplus$ 11B$_{3u}$ $\oplus$ 11B$_{1g}$ $\oplus$ 10B$_{2g}$ $\oplus$ 11B$_{3g}$ $\oplus$ 10A$_{g}$ $\oplus$ 11A$_{u}$
\par The optical modes B$_{1u}$, B$_{2u}$, and B$_{3u}$ are IR active, whereas B$_{1g}$, B$_{2g}$, B$_{3g}$, and A$_{g}$ are Raman active. A$_{u}$ mode vibrations are silent as they do not cause change in polarizability or dipole moment and therefore these modes are neither Raman nor IR active. The calculated optical modes and their vibrational assignment at ambient pressure are given in Table I of the supplementary material. The modes from M96 to M89 and M88 to M81 are due to asymmetric and symmetric stretching of N-H bands, respectively. This is consistent with Dows et al\cite{dows} assignment of the absorption between 2800 and 3200 cm$^{-1}$. The strongest Raman mode frequency 2998.2 cm$^{-1}$ (see figure \ref{fig:Raman}e) in N-H symmetrical stretching region is in excellent agreement with the experimental value of 3000 cm$^{-1}$.\cite{klapotke1} Modes from M80 to M77 originate from N=N=N asymmetric stretching of azide (Type-I $\&$ II) ions are in good accord with measured value of 2030 cm$^{-1}$.\cite{dows} Most of the vibrational modes for NH$_4^+$ ion in AA are assigned based on the vibrational modes of well-known ammonium halides (NH$_4$Cl and NH$_4$Br). Bending motion of NH$_4^+$ ion lies between M76 to M69, and the analogous frequencies in case of NH$_4$Cl\cite{wagner1} and NH$_4$Br\cite{wagner2} are 1712 and 1686 cm$^{-1}$, respectively. Apart from N-H...N short distance from the crystal structure\cite{wu,prince,amorim,millar}, the clear spectroscopic indications of hydrogen bonding in AA\cite{klapotke1,dunsmuir} are due to shift of bending mode frequency above 1400 cm$^{-1}$, torsional and bending motion of NH$_4^+$  ion about 1830 and 2080 cm$^{-1}$, respectively. This is also seen from our present calculations, N-H wagging, rocking and scissoring vibrations arise from M68 to M57 modes and are consistent with measured values of 1400\cite{dows} and 1430\cite{klapotke1} cm$^{-1}$ and also these vibrations are comparable with ammonium halides, 1402 for NH$_4$Cl\cite{wagner1} and 1401 cm$^{-1}$ for NH$_4$Br.\cite{wagner2} The calculated N=N=N symmetric stretching (M56 to M53) modes (doublet is due to presence of Type-I $\&$ II azide ions) of azide ions are consistent with the calculated values for LiN$_3$ (1269.5)\cite{ramesh1} and KN$_3$ (1228.5).\cite{ramesh2} However, the calculated values smaller than the experimental values of 1345\cite{dows} and $\sim$1350\cite{klapotke1} cm$^{-1}$ but the splitting of the doublet is $\sim$ 5.5 cm$^{-1}$ which is in good agreement with experimental value\cite{klapotke1} of $\sim$ 5 cm$^{-1}$.
\par The calculated N=N=N bending and N-H rotational motion of (M52 to M45) modes are consistent with measured values of 664 and 652 cm$^{-1}$.\cite{dows} The modes from M44 to M33 (525.1 to 476.5 cm$^{-1}$) arises from NH$_4^+$ torsional motion and this can be clearly seen from figure \ref{fig:IR}c and the obtained torsional mode frequencies are in good agreement with recent theoretical calculations at ambient pressure.\cite{Wu2} Infrared Spectroscopy\cite{dows} and Inelastic Neutron Scattering \cite{boutin} studies revealed that the torsional mode of NH$_4^+$ ion is located at 420$\pm$20 cm$^{-1}$ and the free rotational motion of NH$_4^+$ ion is prevented by the existence of this torsional motion due to the hydrogen bonding between the nitrogen of NH$_4^+$ ion and the closest nitrogens of azide ions (N-H...N), whereas the torsional motion of Raman bands are very weak, therefore these modes were not observed in the Raman spectroscopic studies at ambient as well as at high pressure.\cite{iqbal,klapotke1} The torsional frequency (in cm$^{-1}$) of ammonium azide and halides decreases in the following order:\cite{dows,satyanarayana} NH$_4$F (523) $\textless$ NH$_4$N$_3$ (420) $\textless$ NH$_4$Cl (359) $\textless$ NH$_4$Br (319) $\textless$ NH$_4$I (297) and this shows that the rotational motion of NH$_4^+$ ion is hindered in both NH$_4$F and NH$_4$N$_3$ than other mentioned ammonium halides due to the strong hydrogen bonding. The mode M32 is purely from the rotation of both NH$_4^+$ and N$_3^-$ ions, whereas M31 and M30 arise from translation motion of NH$_4^+$ ion, which agrees well with the observed\cite{iqbal} Raman mode value of 240 cm$^{-1}$. The modes from M29 to M4 originate from rotational or translation, or a combined rotational and translation motion of NH$_4^+$ and N$_3^-$ ions and the vibrational assignment for each mode is given in Table I of supplementary material. The calculated Raman and IR vibrational modes at ambient pressure are shown in figures \ref{fig:Raman} and \ref{fig:IR}, respectively. As illustrated in figure \ref{fig:Raman}, the calculated Raman spectra at ambient pressure is in close agreement with measured spectra from Raman spectroscopic studies at 0.25 GPa.\cite{klapotke1}

\subsection{IR spectra of AA under pressure}
In general, a structural transformation in materials under compression is experimentally probed by IR and Raman spectroscopy by monitoring the changes in the vibrational spectra under compression.\cite{Landerville2} Therefore, we have calculated the IR spectra under pressure up to 12.0 GPa and it can be divided into five parts as shown in figure \ref{fig:IR}. As illustrated in figures \ref{fig:IR}a and b, as pressure increases the frequency of lattice modes (M4-M52) are found to increase, especially the modes in the Far-IR region (below 300 cm$^{-1}$) which implies hardening of the lattice under pressure. The lattice modes include NH$_4^+$, N$_3^-$ translation, rotational (see figure \ref{fig:IR}a $\&$ b), NH$_4^+$ torsion and N$_3^-$ bending (see figure \ref{fig:IR}c) vibrational modes. The N=N=N asymmetric modes (M77-M80) show a blue shift under compression (see figure \ref{fig:IR}e) whereas the N-H scissoring, rocking (M57-M68) and N-H stretching (M81-M96) modes exhibit a red shift under compression (see figures \ref{fig:IR}d and f). Due to the presence of hydrogen bonding it can be expected that the N-H stretching frequency decreases with increase in pressure\cite{reynolds} and this compression leads to strengthening of hydrogen bonding. Joseph et al\cite{joseph} made an unified explanation for strengthening of hydrogen bonding (X-H...Y) based on red/blue shift of IR frequency and its corresponding intensity. The contraction of hydrogen donor bond (X-H) due to electron affinity of X atom and the associated blue shift and decrease of intensity in IR spectrum or elongation of hydrogen donor bond (X-H) due to attraction between the positive H and electron rich Y atom. As shown in figures \ref{fig:IR}d $\&$ f, the N-H scissoring, rocking and stretching mode frequencies show a red shift and an increase in the intensity of the corresponding vibrational modes up to 4 GPa, which clearly indicates strengthening of the hydrogen bonding under compression below 4.0 GPa. Also the N-H stretching frequencies show a blue-shift between 5-12 GPa which shows weakening of the hydrogen bonding as shown in figure \ref{fig:IR}g. As displayed in figure \ref{fig:IR}f $\&$ g, two N-H asymmetric stretching B$_{2u}$ (2988.6 cm$^{-1}$) and B$_{3u}$ (2989.2 cm$^{-1}$) degenerate modes lift the degeneracy whereas the B$_{1u}$ (3012.4 cm$^{-1}$) and B$_{3u}$ (3015.1 cm$^{-1}$) non-degenerate modes become degenerate under pressure. This is due to strong and weak coupling between N-H band of NH$_4$ and terminal nitrogens of type-I and II azide ions which strengthens and weakens the hydrogen bonding, respectively in the pressure range 0-4 GPa. In addition, we also observe the intensity of asymmetric N-H stretching mode B$_{2u}$ (3001.2 cm$^{-1}$) decreases with pressure and it diminishes between 3-4 GPa. Weakening of hydrogen bonding and disappearance of B$_{2u}$ mode might be the driving force for the structural transition from phase I $\rightarrow$ II as seen in the experiments.\cite{wu,klapotke1}

\subsection{Phonon dispersion curves of AA under high pressure}
In order to get more insight on the dynamical stability of crystalline AA under high pressure, the phonon dispersion curves are calculated at 0 and 6 GPa with 1 $\times$ 2 $\times$ 1 super cell (64 atoms). The computed phonon dispersion curves are plotted along Z-T-Y-S-X and U-R high symmetry directions of the Brillouin zone. As depicted in figure \ref{fig:disp_1}, all the lattice modes ($<$ 700 cm $^{-1}$) shift towards high frequency region except a phonon branch that shows softening (from 38.8 to 35.5 cm$^{-1}$) along the high symmetry S-X-direction when compared at 0 and 6 GPa pressures (see figures \ref{fig:disp_1}a $\&$ b). It is also found that the lattice phonon branches become dispersive at 6 GPa. However, the phonon frequencies are still real along X-direction at 6 GPa which shows the dynamical stability of AA at this pressure. The degenerate N$_3$  bands around $\sim$ 620 cm$^{-1}$ at 0 GPa become non-degenerate modes at 6 GPa (see figures \ref{fig:disp_1} c $\&$ d). As illustrated in figure \ref{fig:disp_2}a, the lowest lying Raman bands correspond to symmetric stretching of Type-I $\&$ II azide ions at 0 GPa which persist even at high pressure 6 GPa, but the Raman measurements\cite{wu,klapotke1} have only one stretching band in this region above 3 GPa which is due to the fact that the two azides occupy equivalent Wyckoff positions in the high pressure phase (see \ref{fig:disp_2}a $\&$ b). The present and recent theoretical calculations\cite{abstract} are unsuccessful in reproducing the Raman spectral evolution under high pressure as observed in the experiments.\cite{wu,klapotke1} The middle and top phonon bands correspond to N-H bending modes which show a red shift when compared at 0 and 6 GPa pressures as displayed in figures \ref{fig:disp_2}a $\&$ b. The N=N=N asymmetric phonon bands show a blue-shift with pressure as depicted in figures \ref{fig:disp_2}c $\&$ d. The high frequency N-H symmetric and asymmetric stretching bands show a red shift as presented in figures \ref{fig:disp_2}e $\&$ f. However, the high frequency modes show a red shift up to 4 GPa and later on they exhibit blue-shift as shown in figure \ref{fig:IR}g. We also observe that there is a deviation in the calculated phonon frequencies using NC (see figures \ref{fig:Raman} and \ref{fig:IR}) and PAW (see figures \ref{fig:disp_1} and \ref{fig:disp_2}) pseudo potentials, especially for N=N=N symmetric and asymmetric stretching phonon modes. Overall, we observe that all the vibrational modes are real for the ambient phase (I) of AA at 0 and 6 GPa. This implies that AA is found to be dynamically stable at ambient as well as at high pressure (6 GPa) as depicted in figures \ref{fig:disp_1} and \ref{fig:disp_2} and also suggest that the structural transition might be due to the weakening of the hydrogen bonding under high pressure.

\section{Conclusions}
In summary, we made a detailed study of phase stability, structural and vibrational properties of AA under hydrostatic compression up to 50 GPa with dispersion correction methods. The calculated ground state parameters at ambient as well as at high pressure using vdW-TS and TS-SCS methods are in good agreement with experimental data and this study shows the importance of semi-empirical dispersion correction methods to capture dispersive interactions in molecular solids. The calculated compression curves are consistent with the trend followed by the elastic constants as well as with high pressure experimental data. Softening of the shear elastic moduli is suggestive of mechanical instability of AA under high pressure. AA is found to be the thermodynamic ground state of N$_4$H$_4$ compounds and it undergoes a series of phase transitions from AA $\rightarrow$ TTZ at around $\sim$ 39-43 GPa and TTZ $\rightarrow$ HNS-1 at 80-90 GPa under the studied temperature range of 0-650 K which are consistent with the recent theoretical calculations. The anisotropic compressibility of a and c lattice constants suggests that the ambient phase of AA undergoes a transition to tetragonal phase, and then to a less anisotropic low  symmetry crystalline phase above 22 GPa. We predict a rapid decrease in the angle that Type-II azides make with c-axis ($\theta$) till 12 GPa and then an increase up on further compression. In addition to that we have also calculated the zone centre phonon frequencies, IR spectra using DFPT at ambient and under high pressure. The calculated vibrational frequencies at ambient pressure and their complete vibrational assignment are consistent with experimental observations. AA is found to be dynamically stable at ambient pressure and no soft phonon mode is observed beyond the experimental transition pressure from phonon dispersion curves. Also the calculated IR spectra show that the N-H stretching frequencies get a red and blue-shift below and above 4 GPa, respectively. The intensity of the B$_{2u}$ mode is found to diminish gradually and the weak coupling between NH$_4$ and N$_3$ ions makes B$_{1u}$ and B$_{3u}$ modes to be degenerate with progression of pressure up to 4 GPa which causes weakening of hydrogen bonding and these effects lead to a structural phase transition in AA around 4 GPa and this is in good agreement with the experimental observations.

\section{Acknowledgments}
NYK and GV would like to thank Defense Research and Development Organization (DRDO) through ACRHEM for the financial support under grant No. DRDO/02/0201/2011/00060:ACREHM-PHASE-II, and the CMSD, University of Hyderabad, for providing computational facilities. NYK thanks Prof. C. S. Sunandana, School of Physics, University of Hyderabad, for critical reading of the manuscript. PA would like to acknowledge B. K. Panigrahi for his support.\\
$^*$\emph{Author for Correspondence, E-mail: gvaithee@gmail.com}

\clearpage

\begin{figure}[h]
\centering
{\subfigure[]{\includegraphics[height = 2.8in, width=2.0in]{Fig5a.eps}}} \hspace{0.1in}
{\subfigure[]{\includegraphics[height = 2.8in, width=3.5in]{Fig5b.eps}}}
{\subfigure[]{\includegraphics[height = 2.8in, width=4.0in]{Fig5c.eps}}}  \hspace{0.1in}
{\subfigure[]{\includegraphics[height = 2.3in, width=3.0in]{Fig5d.eps}}}
\caption{(Colour online) Calculated (a, b) lattice constants a, b and c; (c) relative compressibility of unit cell parameters a, b and c; (d) volume of AA as a function of pressure up to 2.5 GPa (except figure 5(b) which is up to 50 GPa). The experimental data is taken from Refs.\cite{wu,millar,laura}}
\label{fig:lattice}
\end{figure}


\begin{figure}[h]
\centering
{\subfigure[]{\includegraphics[height = 2.8in, width=3.5in]{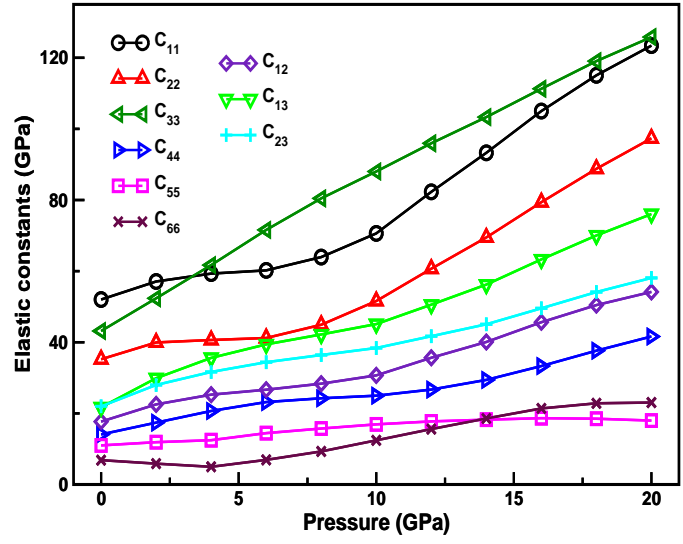}}}  \vspace{0.2in}
{\subfigure[]{\includegraphics[height = 2.8in, width=3.5in]{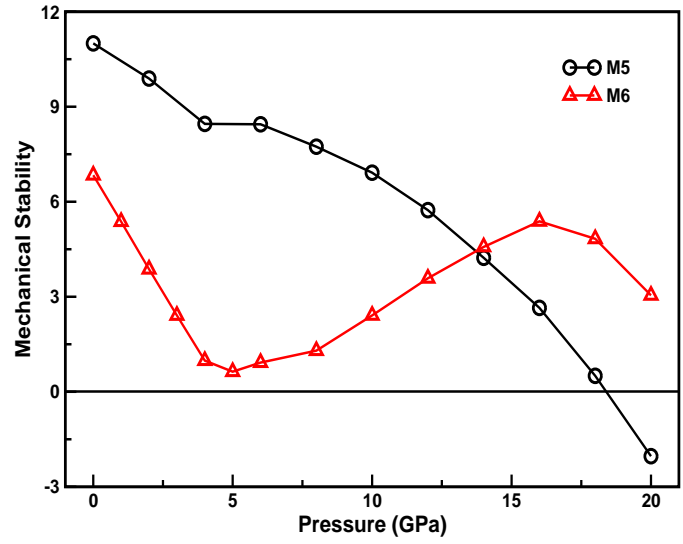}}}
\caption{(Colour online) Calculated (a) elastic constants (C$_{ij}$) (b) mechanical stability criteria of AA as a function of pressure up to 20 GPa.}
\label{fig:elastic}
\end{figure}

\begin{figure}[h]
\centering
{\subfigure[]{\includegraphics[height = 2.5in, width=3.5in]{Fig7a.eps}}}  \hspace{0.1in}
{\subfigure[]{\includegraphics[height = 2.5in, width=3.5in]{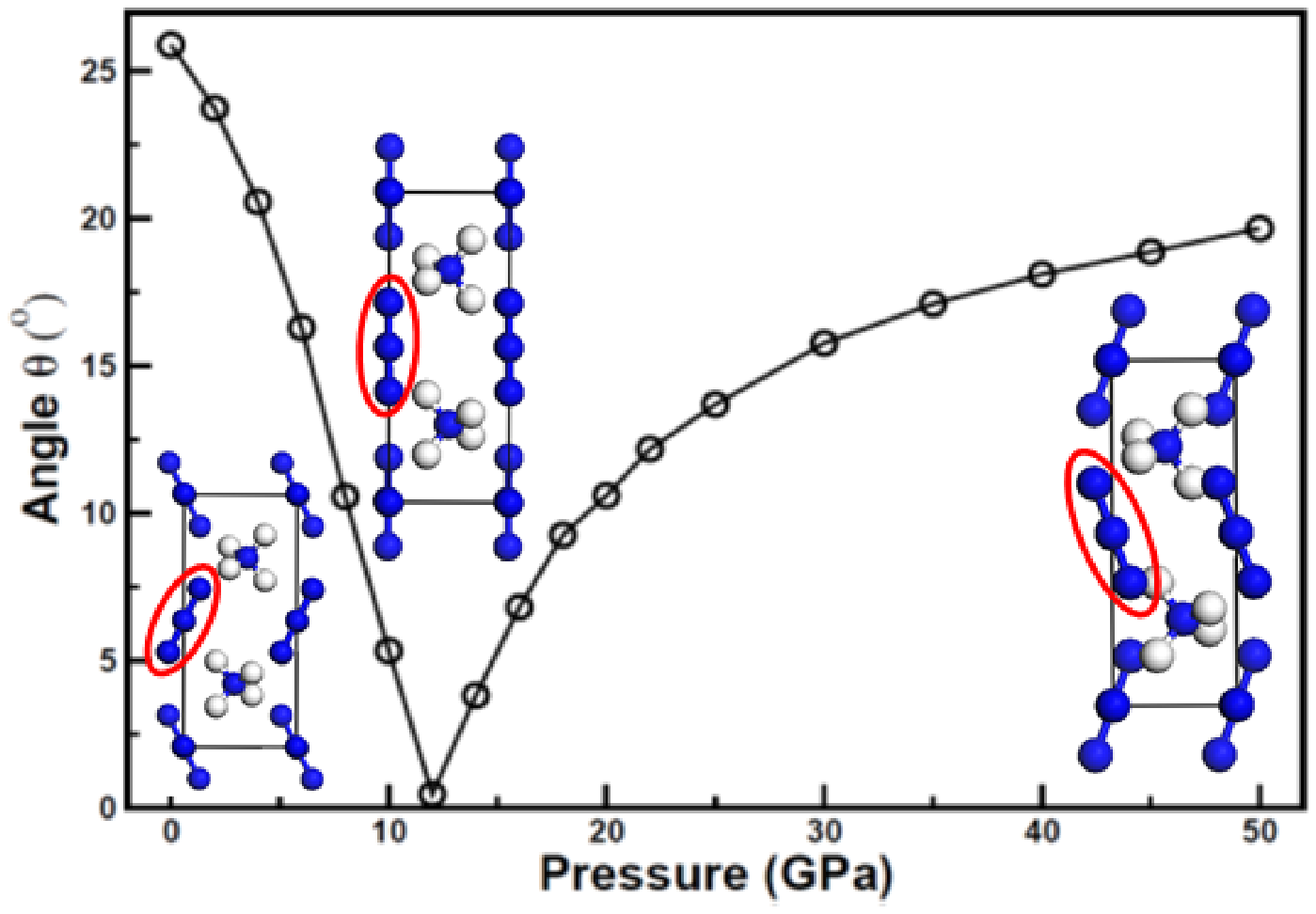}}}
\caption{(Colour online) Calculated angle $\theta$ that a Type-II azide ion makes with c-axis (see figure \ref{fig:AA}) in AA up to (a) 0-2.5 GPa and (b) 0-50 GPa. The optimized structures of AA are given in inset at 0, 12 and 50 GPa pressures (from left to right) in order to show change in the orientation of azide ions with c-axis in figure (b) as highlighted in ellipse for one Type-II azide and it is also applicable for Type-II azides which are aligned along c-axis. The experimental data is taken from Ref.\cite{wu}}
\label{fig:beta}
\end{figure}

\begin{figure}[h]
\centering
\includegraphics[height = 3.0in, width=4.5in]{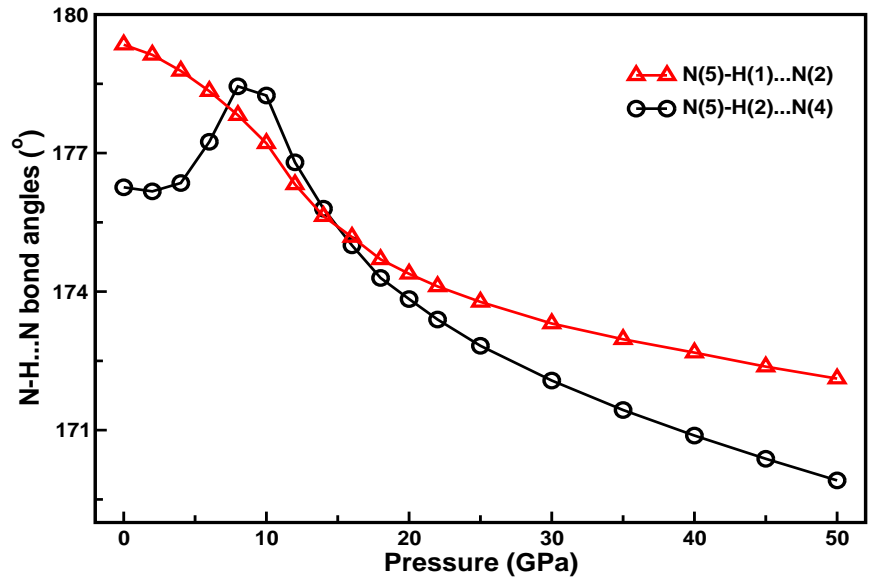}
\caption{(Colour online) Calculated hydrogen bond angles (N-H...N) as a function of pressure up to 50 GPa. Atom labels are given in figure \ref{fig:AA}.}
\label{fig:BA_H}
\end{figure}

\begin{figure}[h]
\centering
{\subfigure[]{\includegraphics[height = 2.5in, width=2.5in]{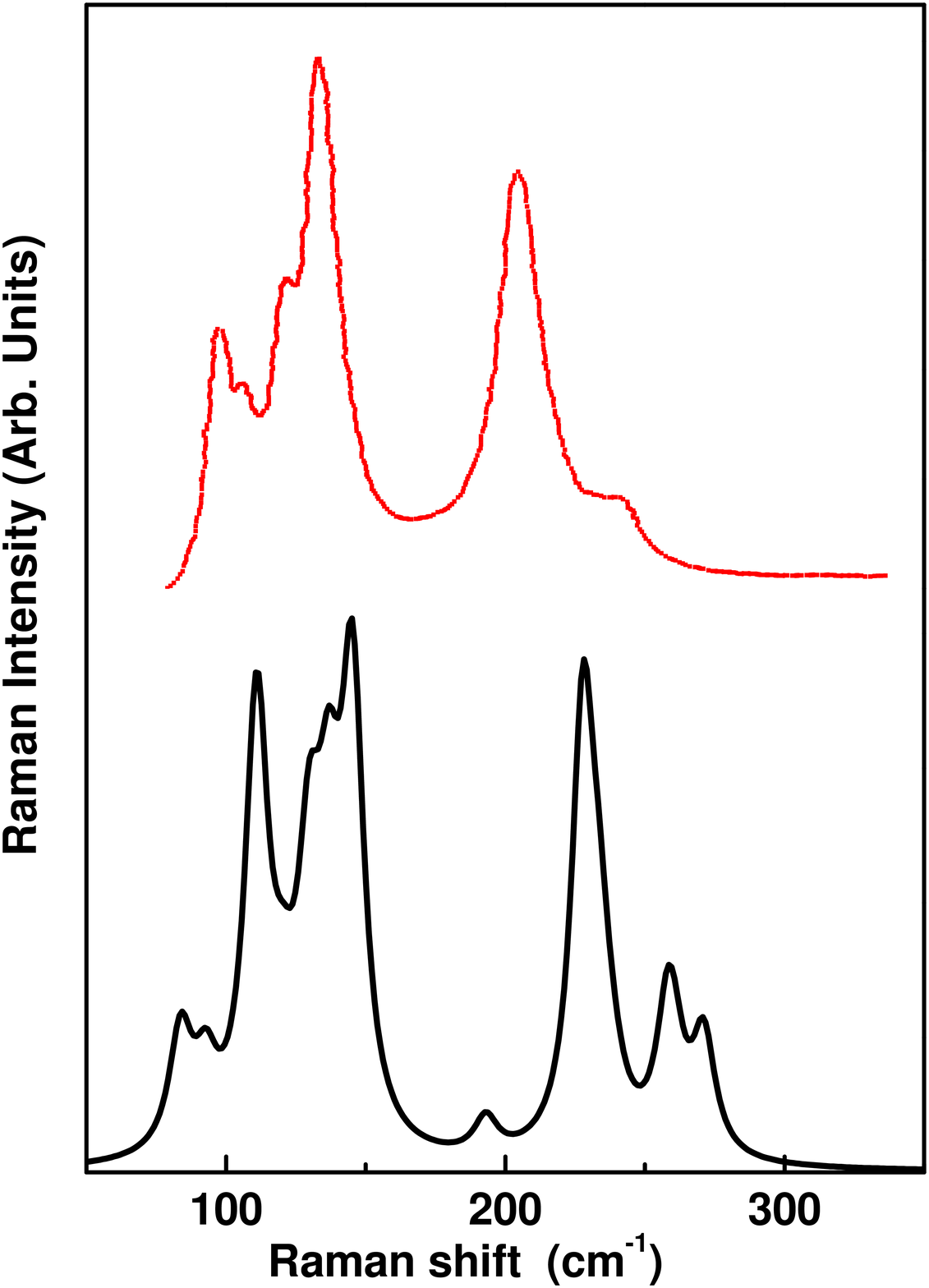}}}  \hspace{0.1in}
{\subfigure[]{\includegraphics[height = 2.8in, width=2.8in]{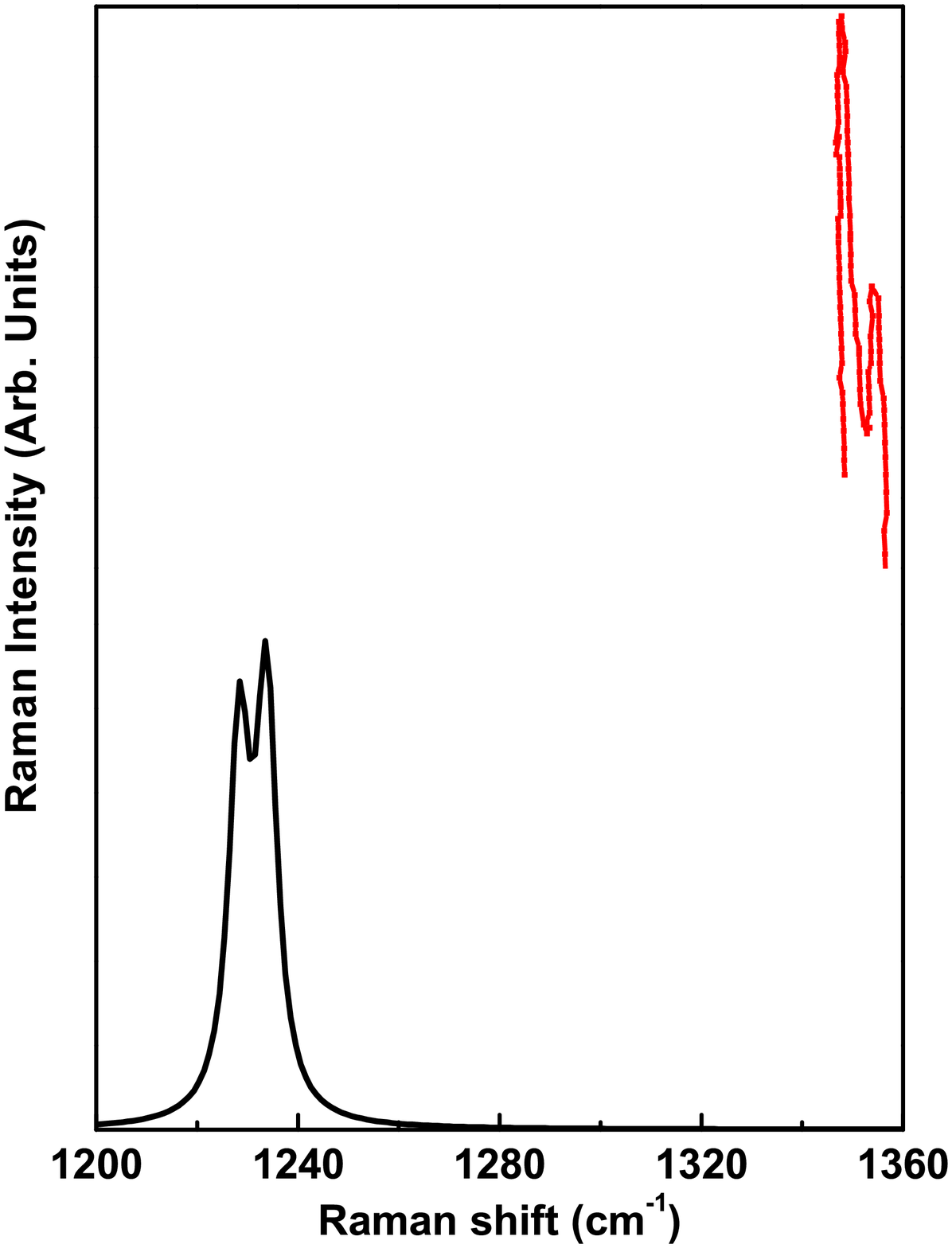}}}  \hspace{0.1in}
{\subfigure[]{\includegraphics[height = 2.6in, width=2.8in]{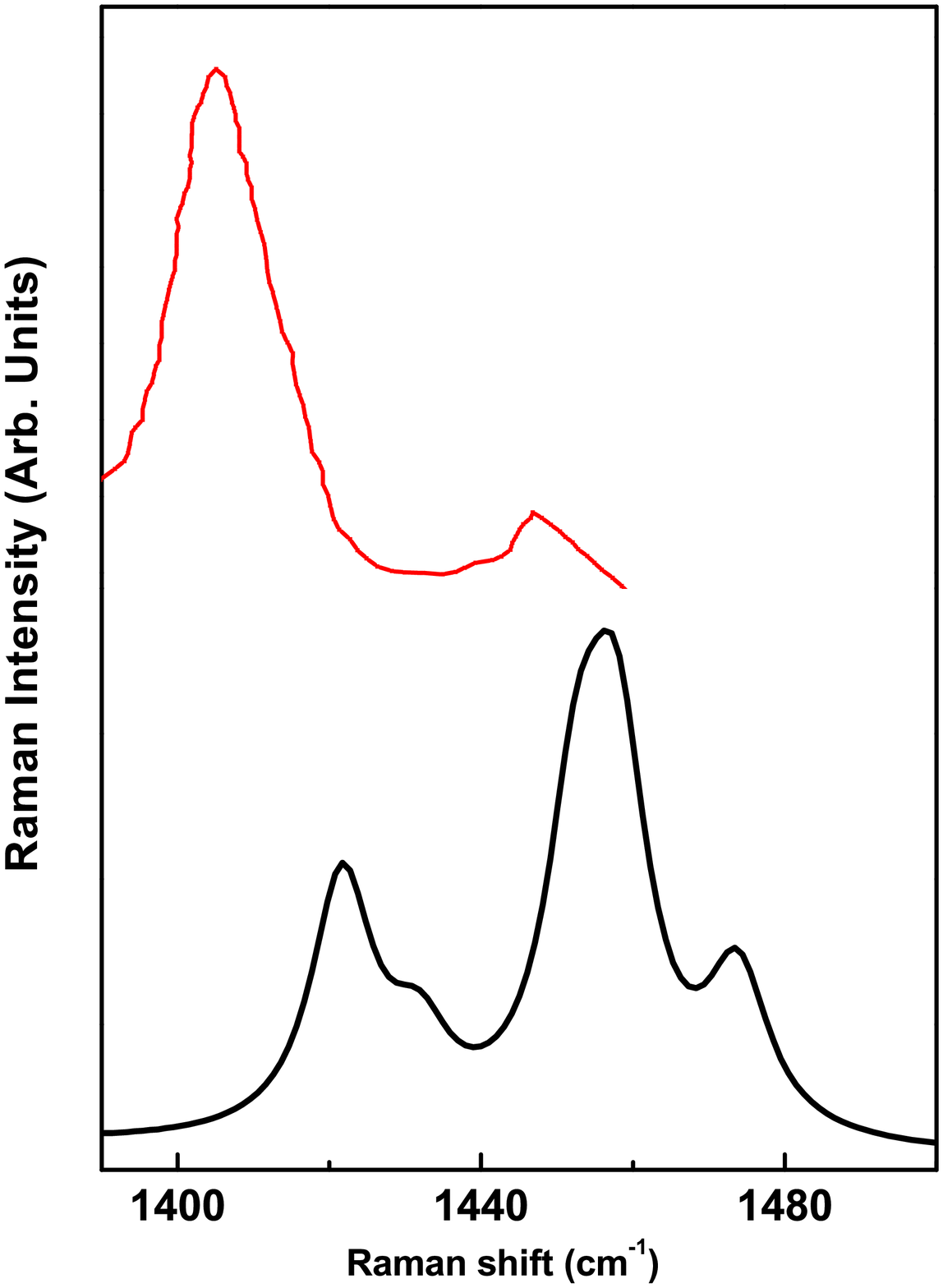}}}  \hspace{0.1in}
{\subfigure[]{\includegraphics[height = 2.6in, width=2.8in]{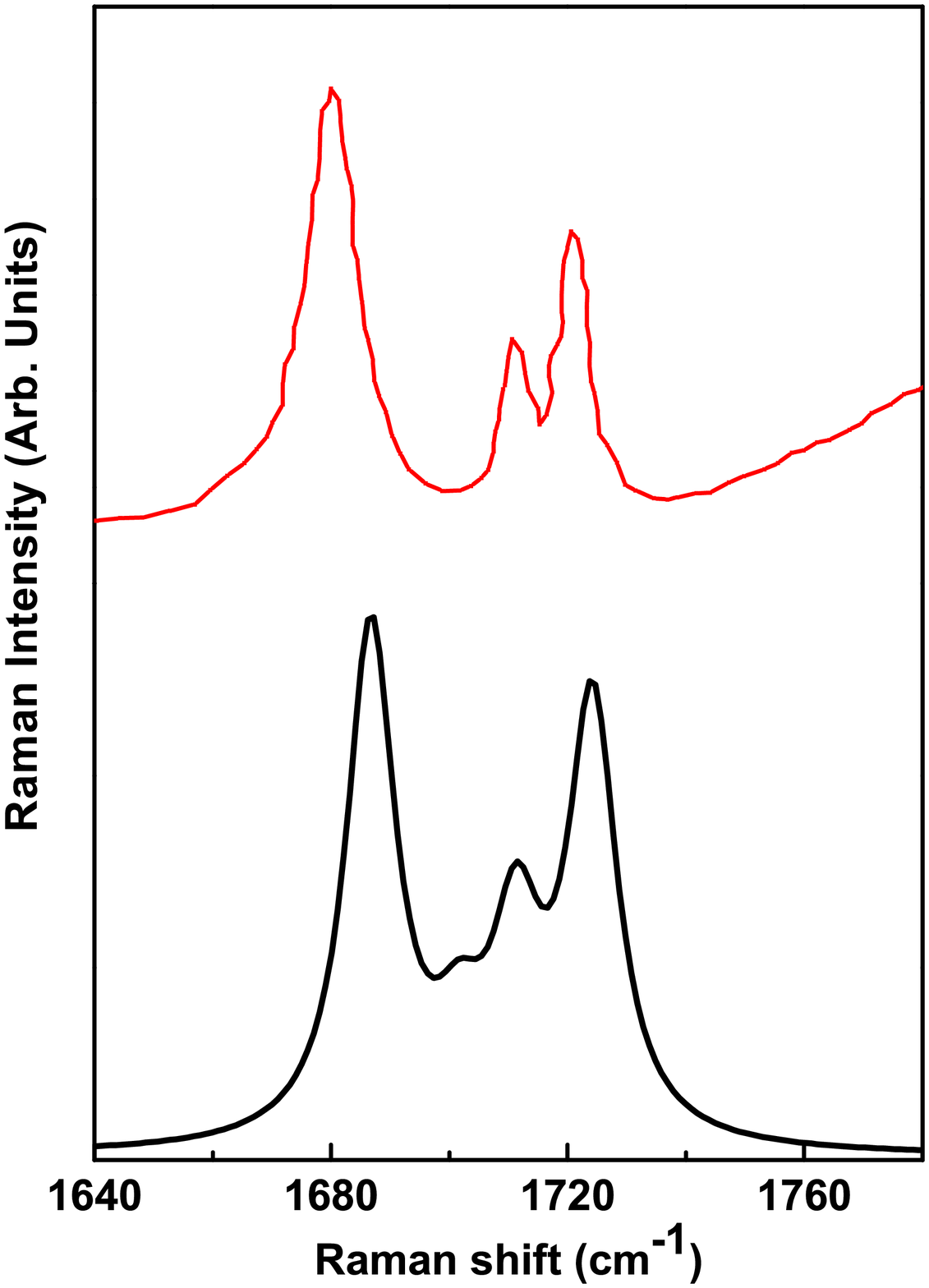}}}  \hspace{0.1in}
{\subfigure[]{\includegraphics[height = 2.8in, width=2.5in]{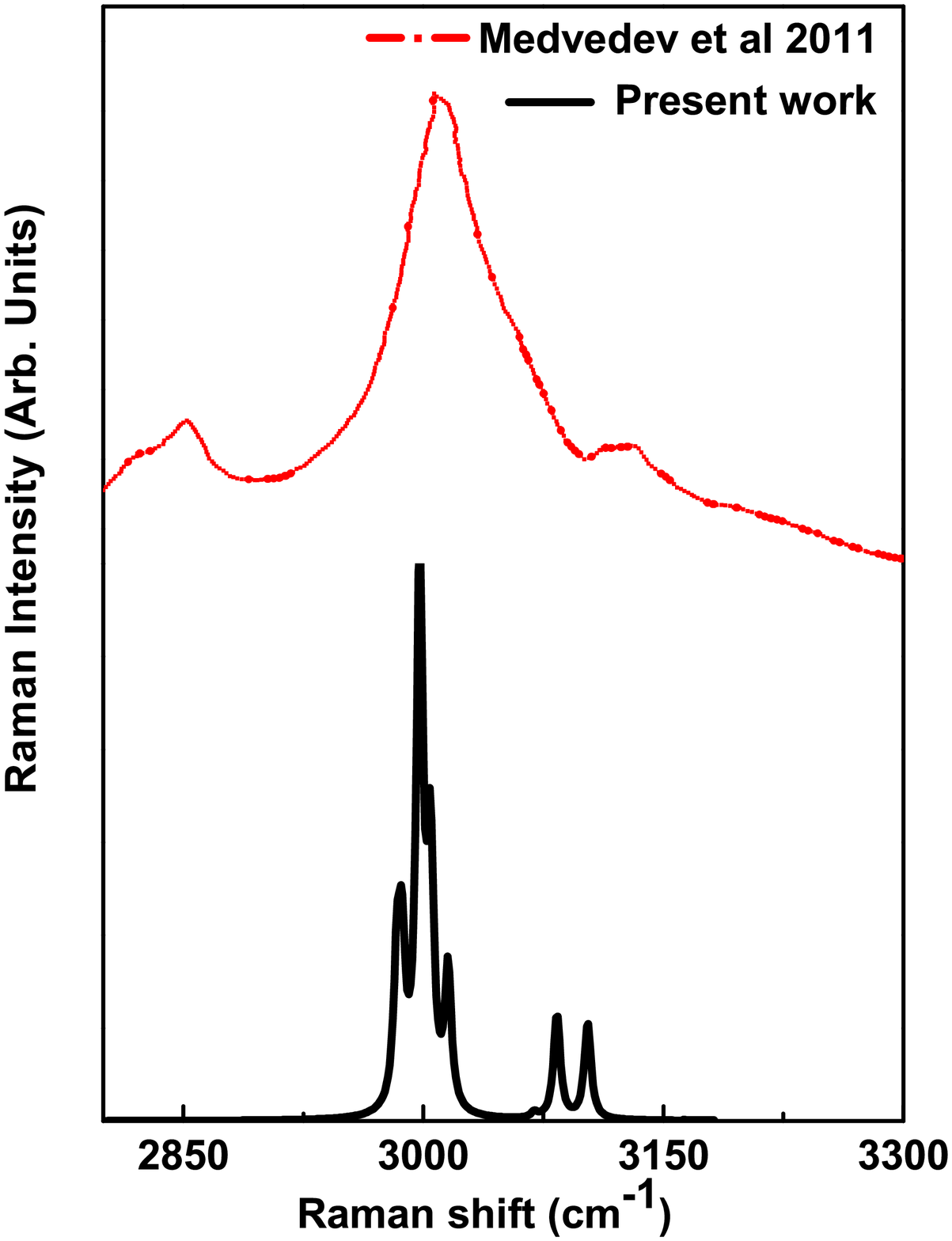}}}  \hspace{0.1in}
\caption{(Colour online) Calculated (bottom, solid black line) and experimental (top, solid red line) Raman spectra of AA at ambient pressure. The experimental data is taken from Ref.\cite{klapotke1}}
\label{fig:Raman}
\end{figure}


\begin{figure}[h]
\centering
{\subfigure[]{\includegraphics[height = 2.2in, width=1.8in]{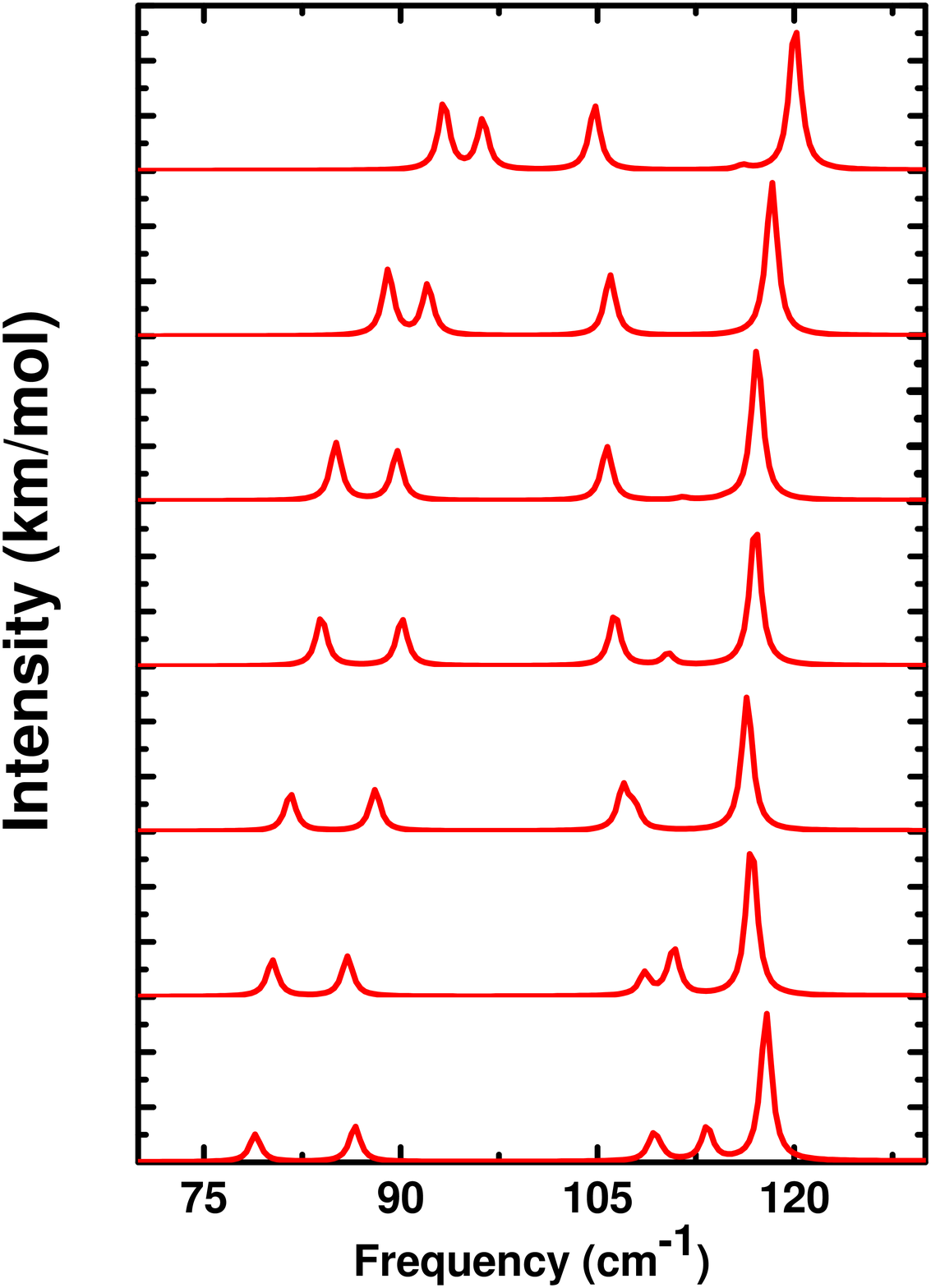}}} \hspace{0.1in}
{\subfigure[]{\includegraphics[height = 2.2in, width=1.8in]{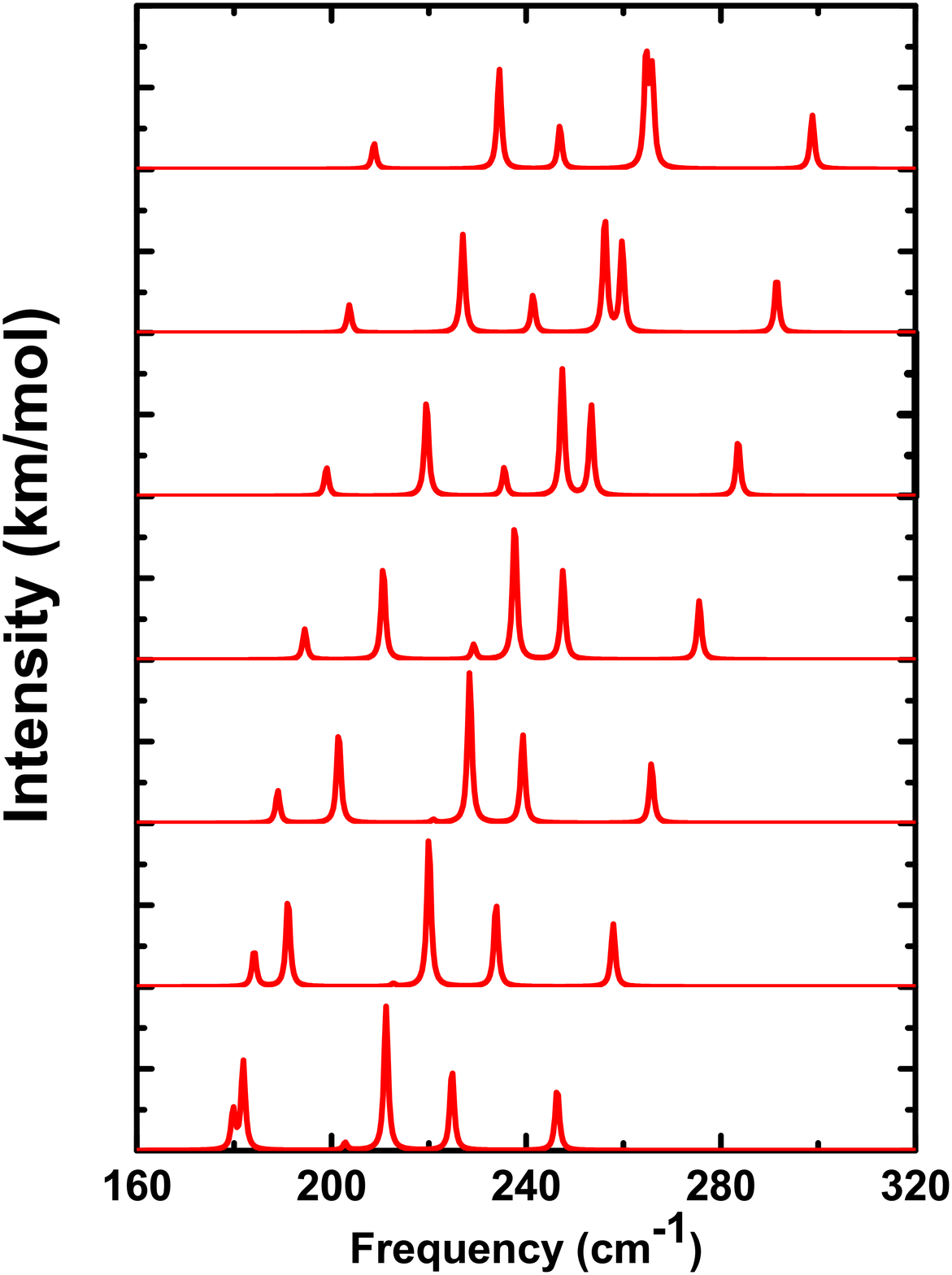}}} \hspace{0.1in}
{\subfigure[]{\includegraphics[height = 2.2in, width=1.8in]{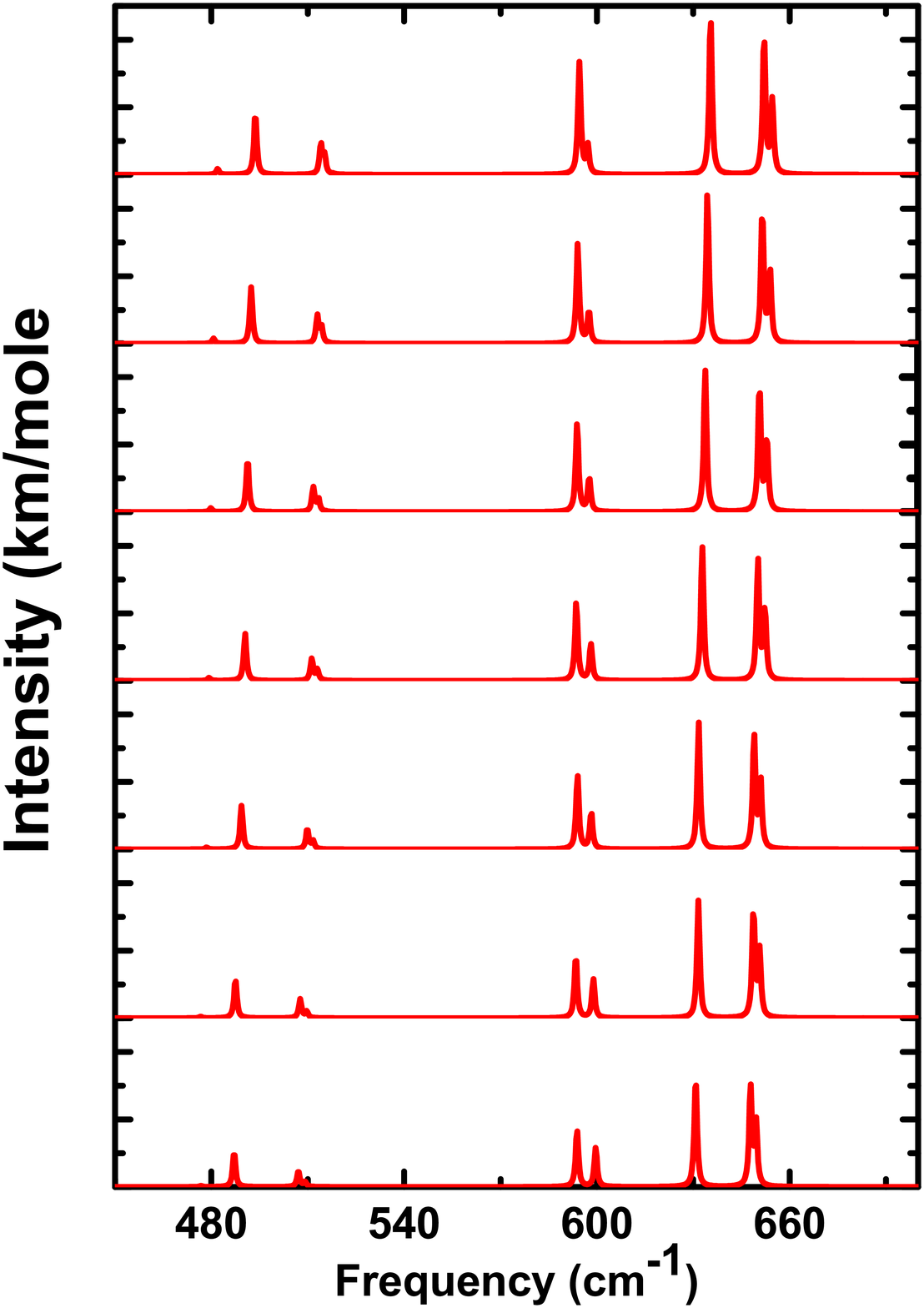}}}  \hspace{0.1in}
{\subfigure[]{\includegraphics[height = 2.2in, width=1.8in]{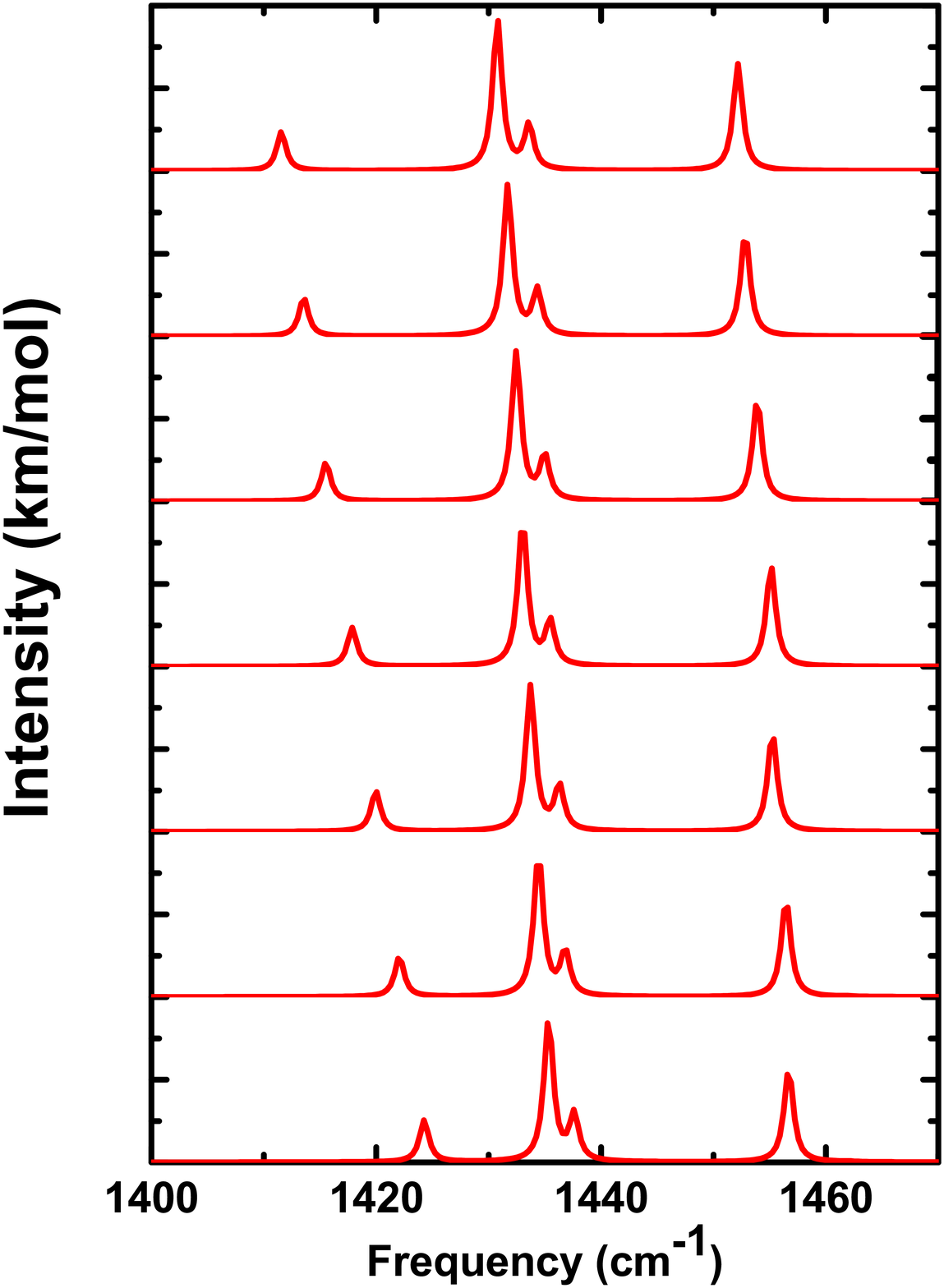}}} \hspace{0.1in}
{\subfigure[]{\includegraphics[height = 2.2in, width=1.8in]{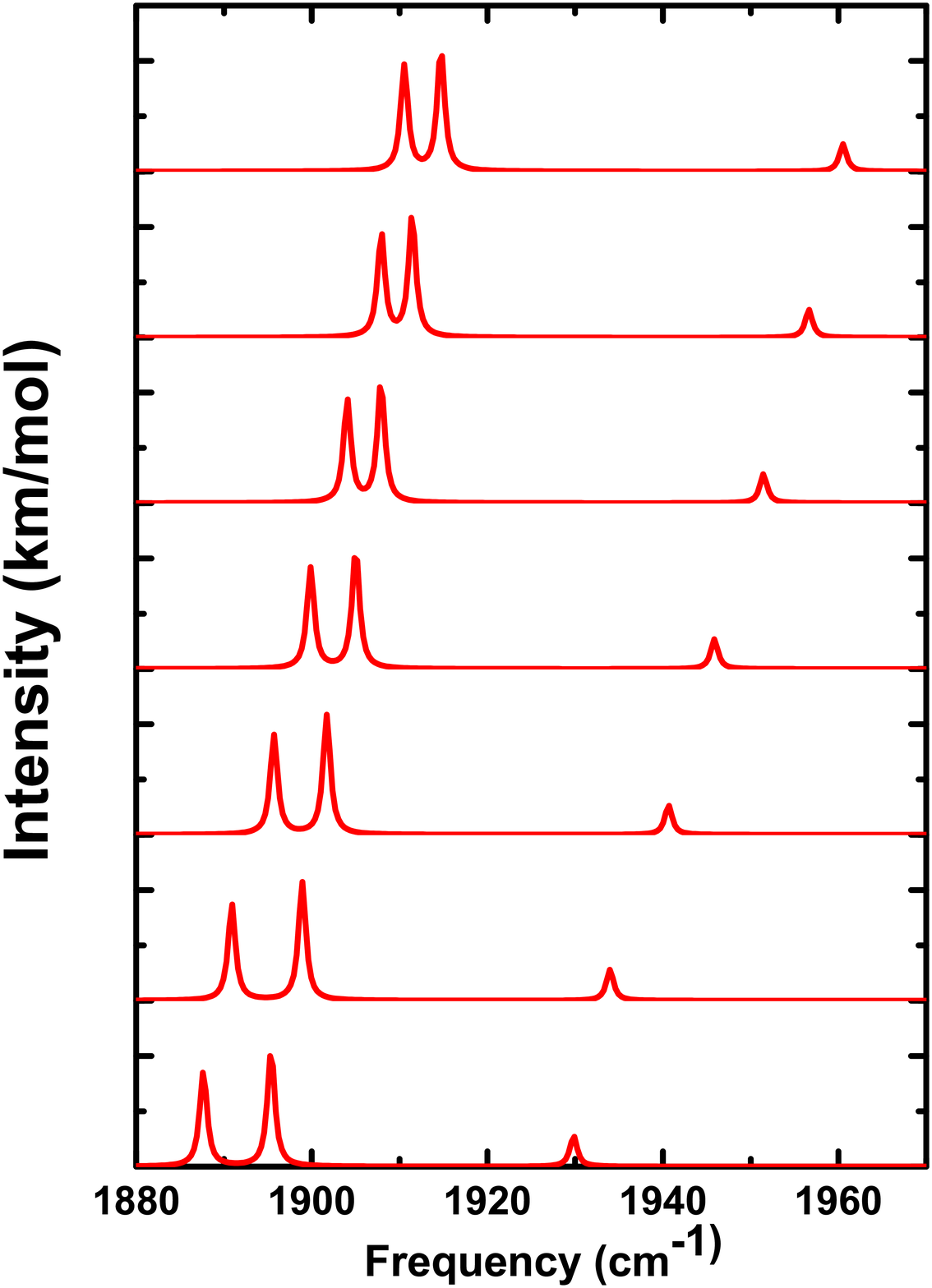}}}  \hspace{0.1in}
{\subfigure[]{\includegraphics[height = 2.2in, width=1.9in]{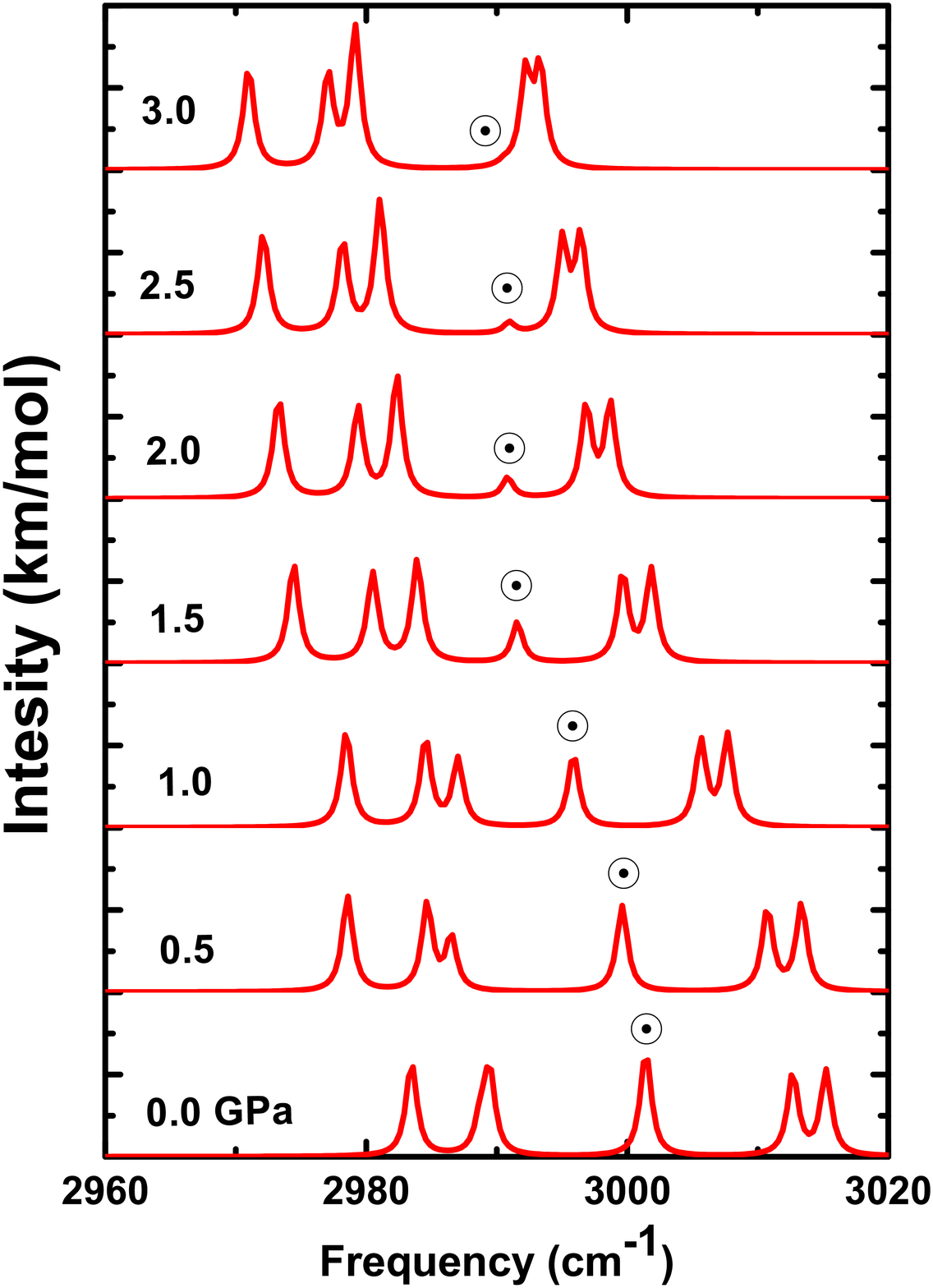}}} \hspace{0.1in}
{\subfigure[]{\includegraphics[height = 2.2in, width=1.8in]{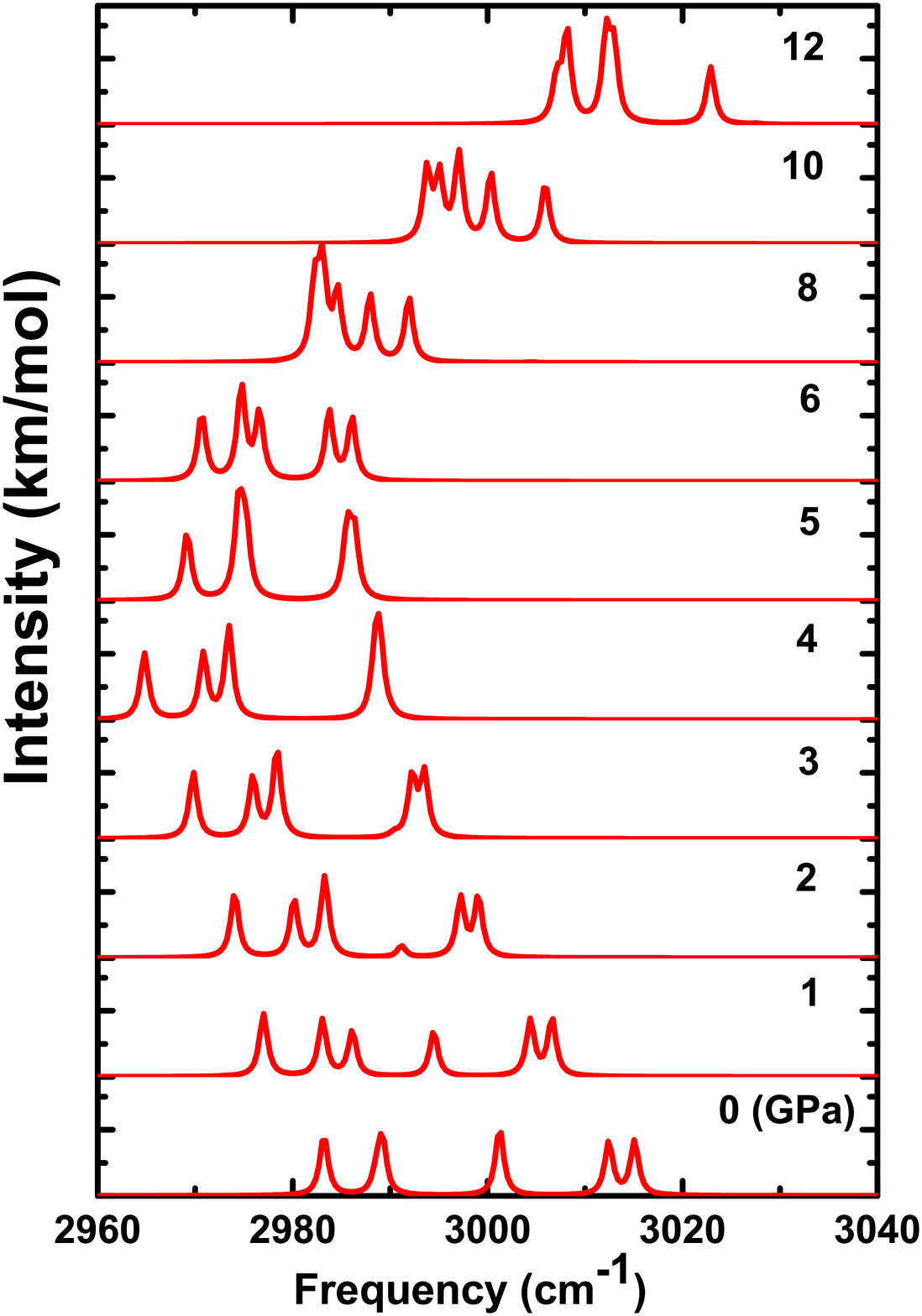}}}
\caption{(Colour online) Calculated IR spectra (a, b) lattice modes (c) torsional and bending modes of NH$_4$ and N$_3$ ions (d) N-H wagging, rocking and scissoring modes of  NH$_4$ ions (e) N$_3$ asymmetric modes (f, g) N-H symmetric and asymmetric stretching modes of AA as a function of pressure. Calculated lattice, N-H torsional + N$_3$ bending and azide stretching modes show blue shift whereas N-H bending and stretching modes show red shift with pressure. In particular the N-H stretching modes show red shift up to 4 GPa thereafter show blue-shift between 5-12 GPa.}
\label{fig:IR}
\end{figure}


\begin{figure}[h]
\centering
{\subfigure[]{\includegraphics[height = 2.5in, width=3.0in]{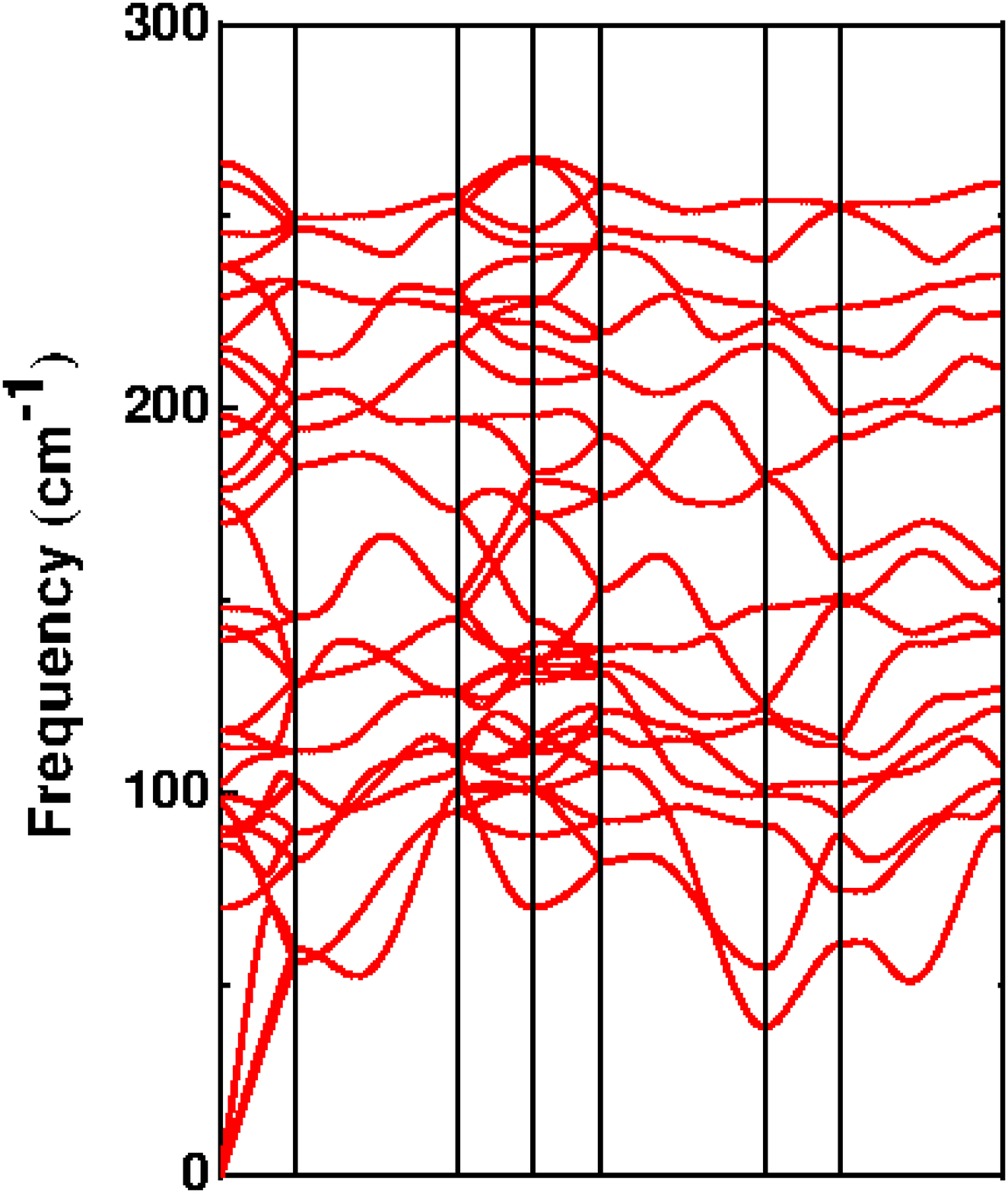}}}
{\subfigure[]{\includegraphics[height = 2.5in, width=3.0in]{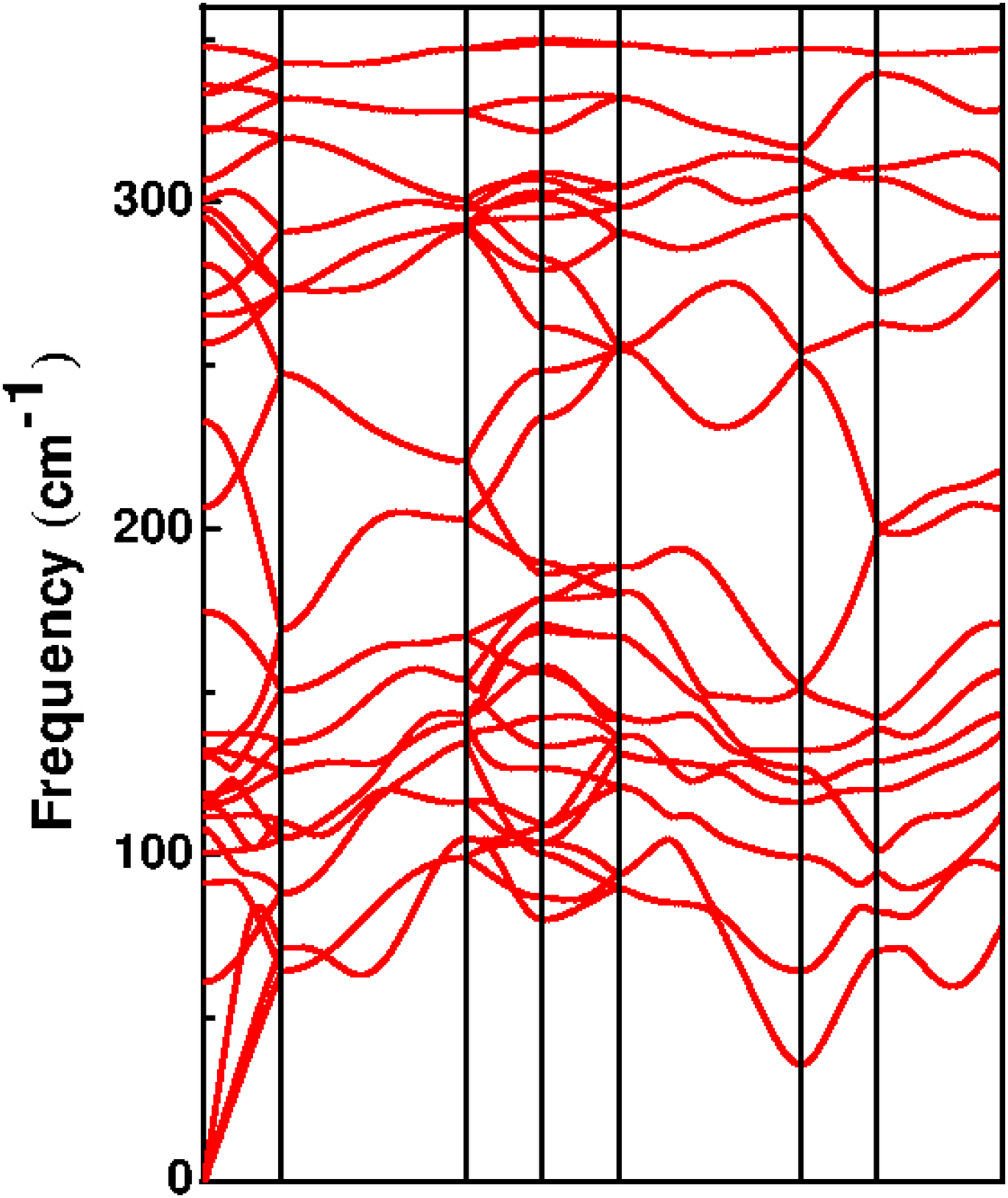}}} 
{\subfigure[]{\includegraphics[height = 2.5in, width=3.0in]{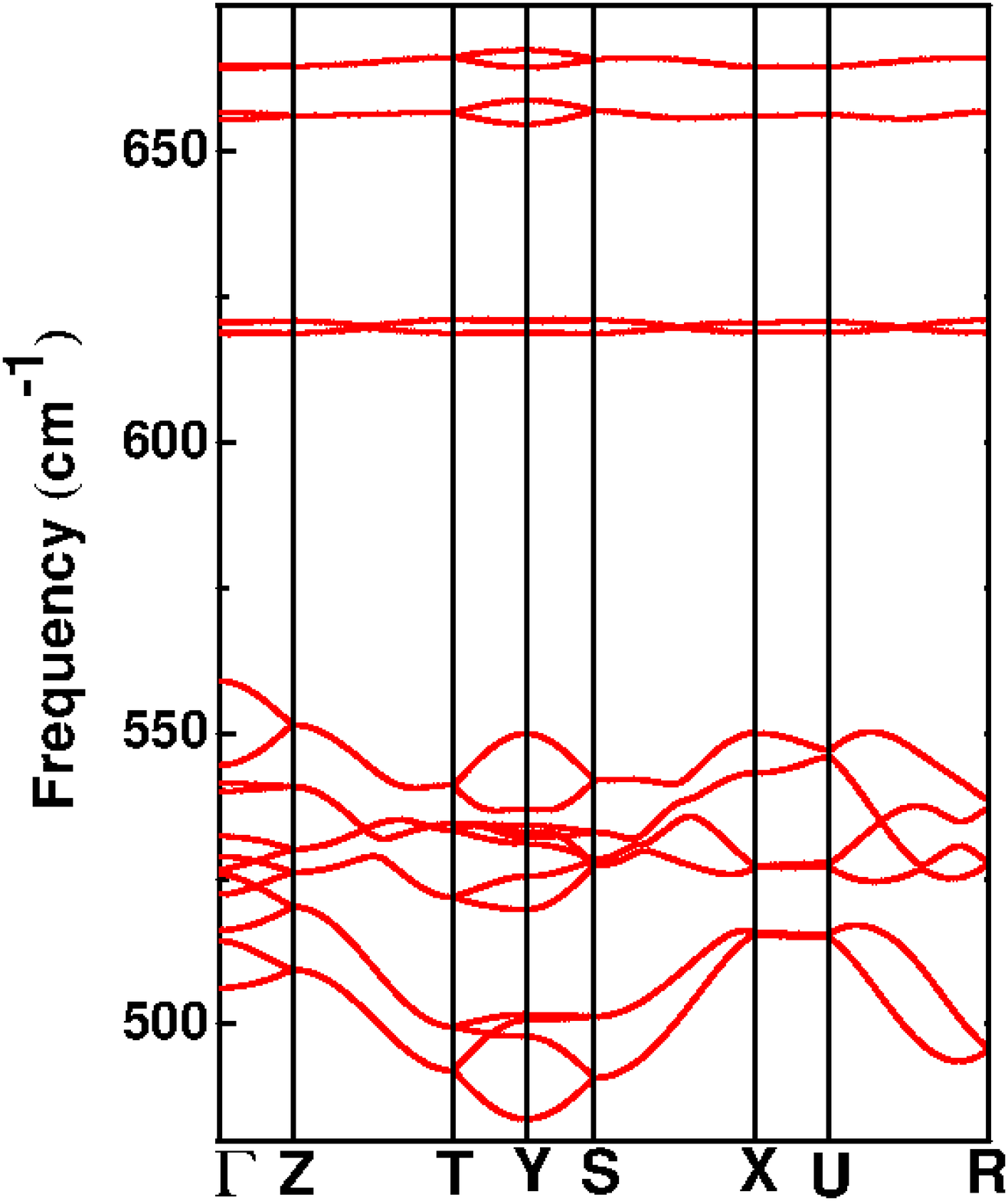}}}
{\subfigure[]{\includegraphics[height = 2.5in, width=3.0in]{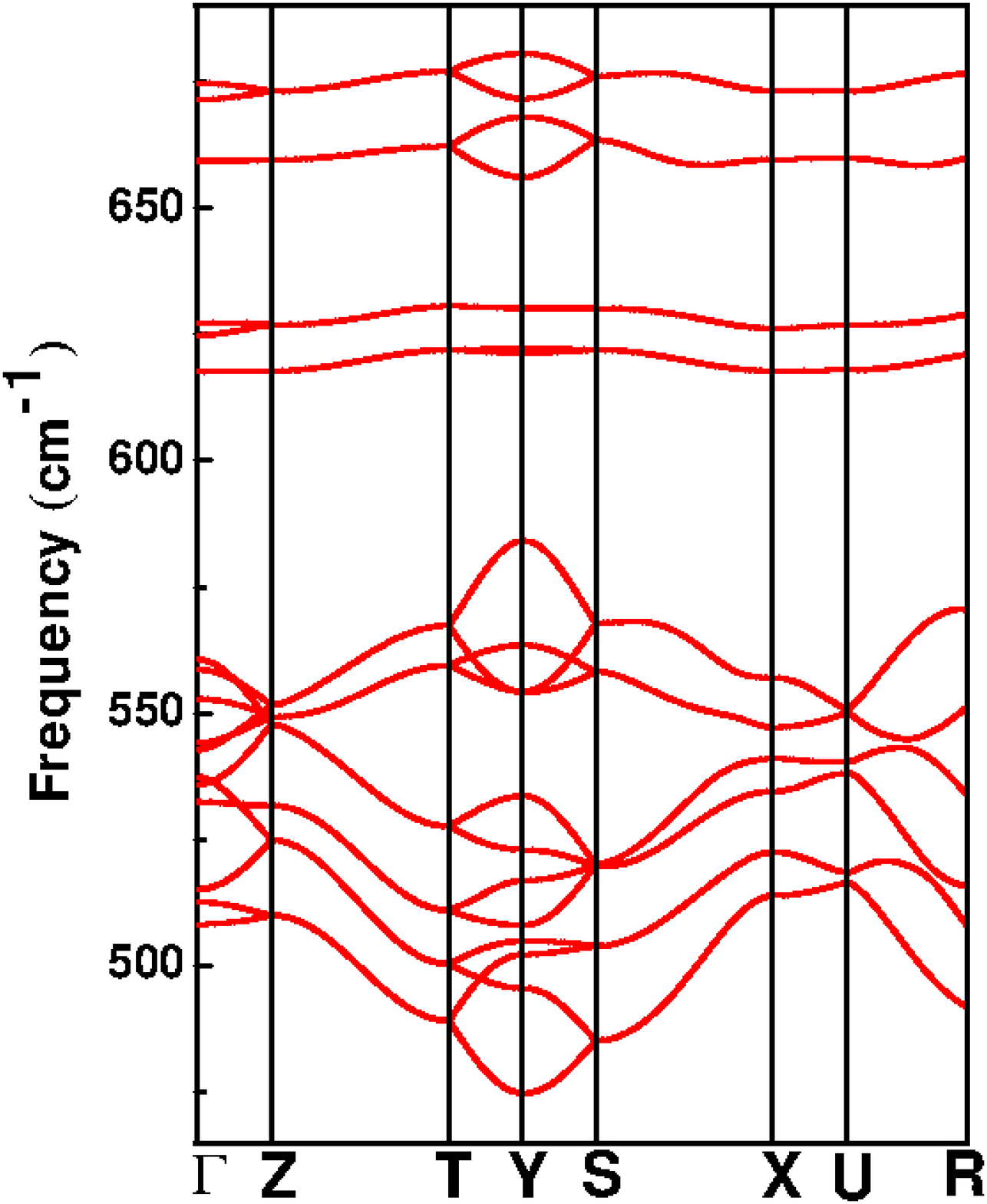}}}
\caption{(Colour online) Calculated phonon dispersion curves in the low frequency region ($\textless$ 700 cm$^{-1}$) of AA at 0 GPa (left) as well as at 6 GPa (right) along high symmetry directions $\Gamma$ (0.0, 0.0, 0.0) $\rightarrow$ Z (0.0, 0.0, 0.5) $\rightarrow$ T (0.0 0.5, 0.5) $\rightarrow$ Y (0.0 0.5 0.0) $\rightarrow$ S (0.5, 0.5, 0.0) $\rightarrow$ X (0.5, 0.0, 0.0) $\rightarrow$ U (0.5, 0.0, 0.5) $\rightarrow$ R (0.5, 0.5, 0.5) in the Brillouin zone.}
\label{fig:disp_1}
\end{figure}

\begin{figure}[h]
\centering
{\subfigure[]{\includegraphics[height = 2.5in, width=3.0in]{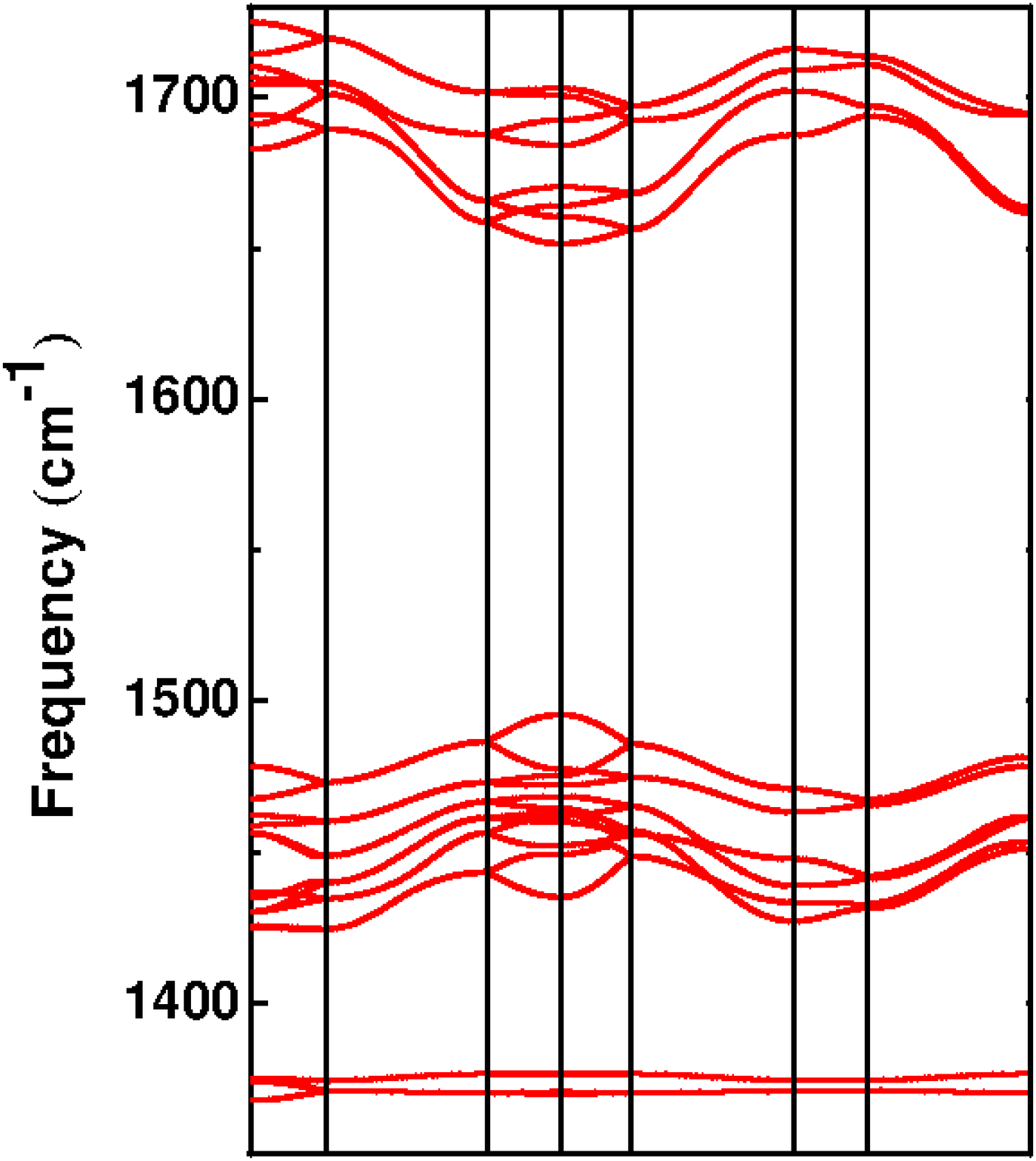}}}
{\subfigure[]{\includegraphics[height = 2.5in, width=3.0in]{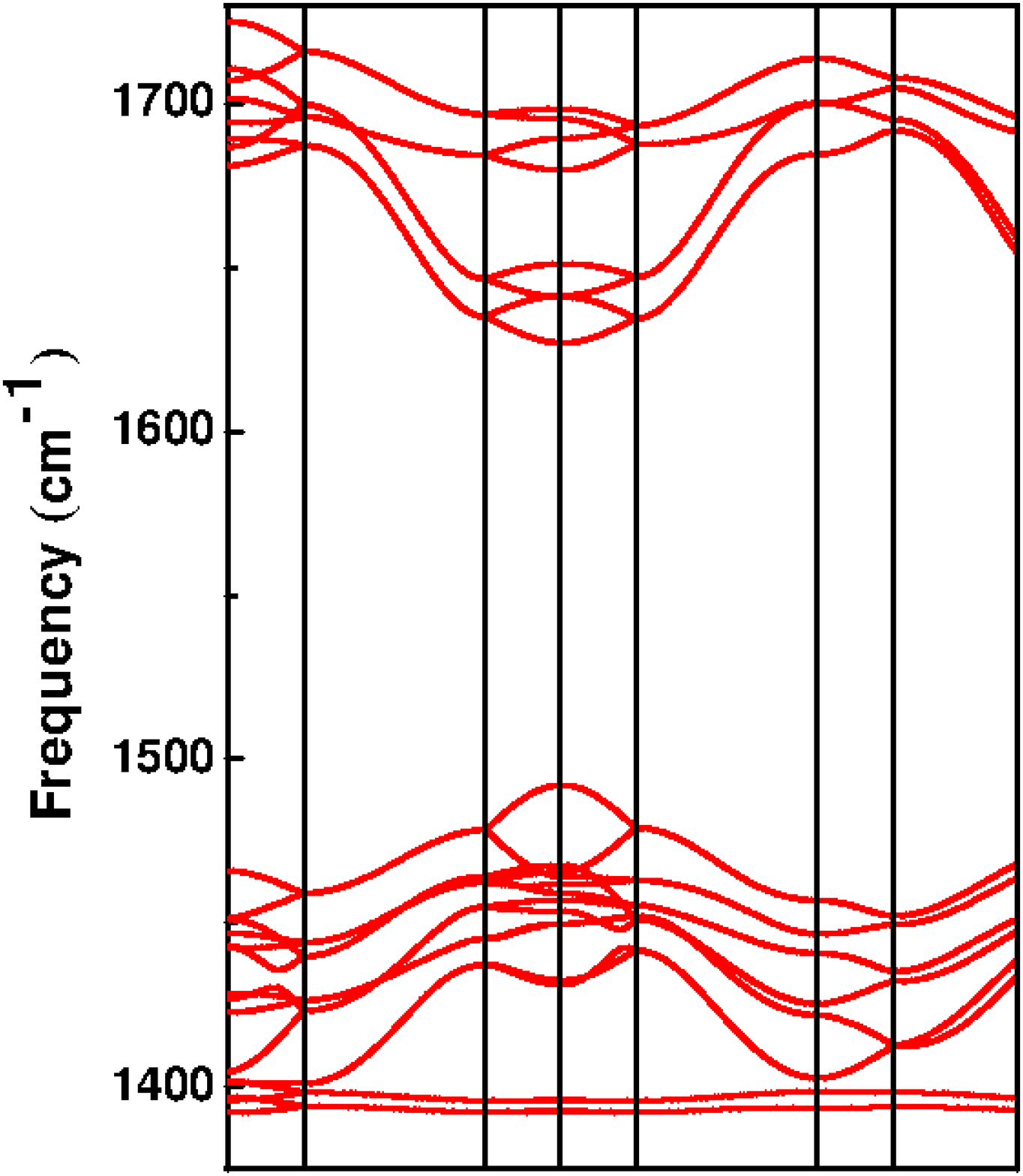}}} 
{\subfigure[]{\includegraphics[height = 2.5in, width=3.0in]{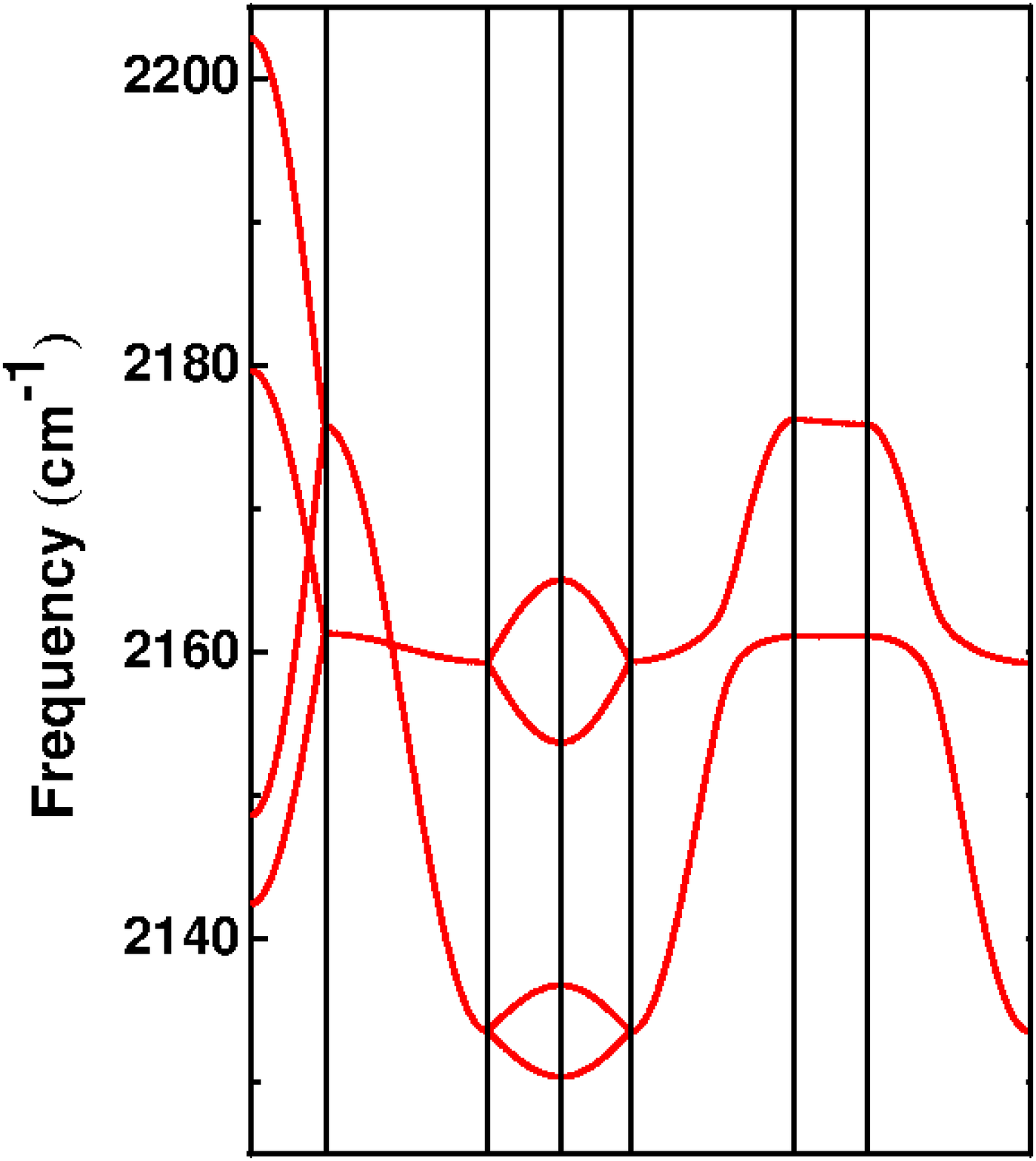}}}
{\subfigure[]{\includegraphics[height = 2.5in, width=3.0in]{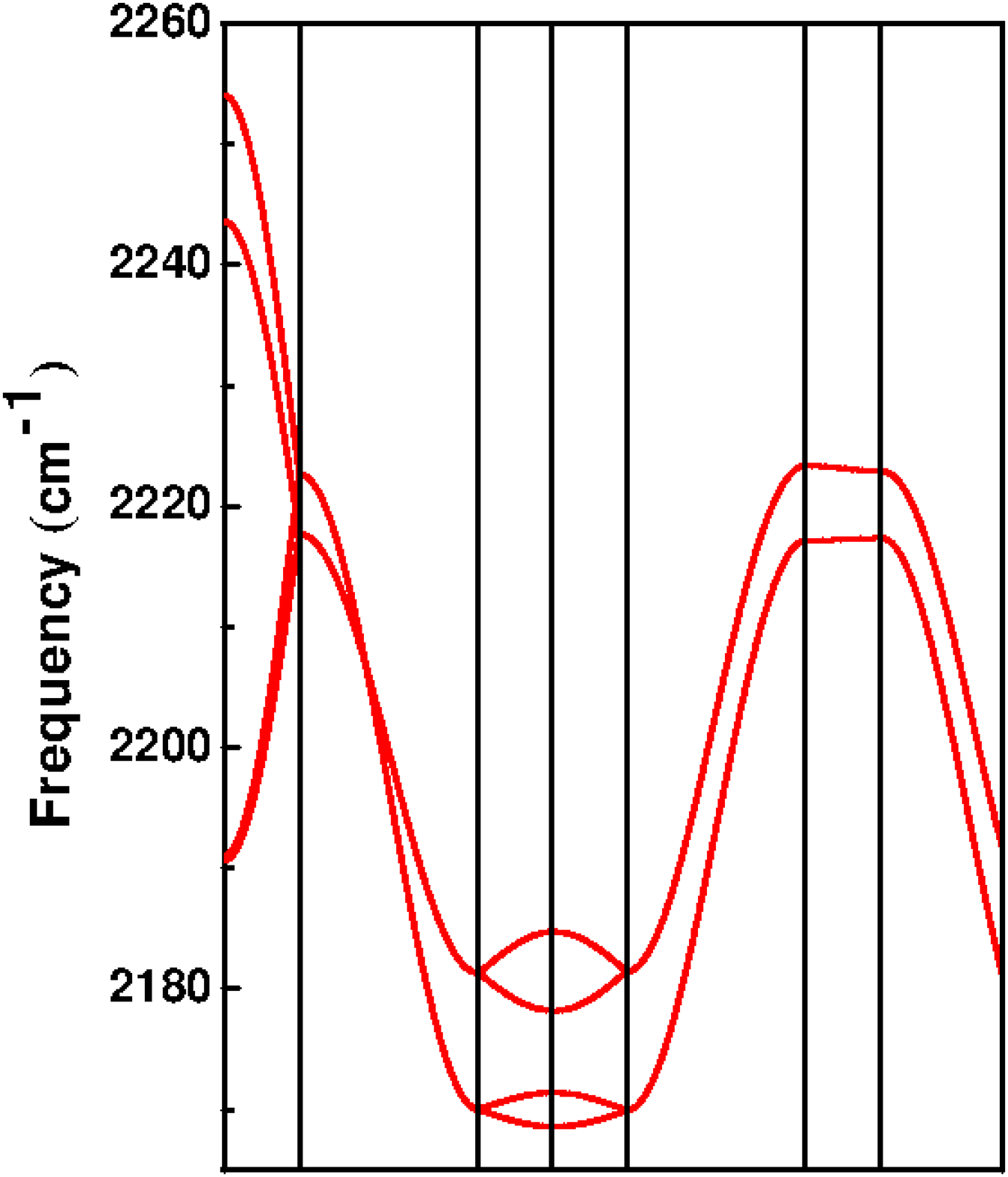}}}
{\subfigure[]{\includegraphics[height = 2.5in, width=3.0in]{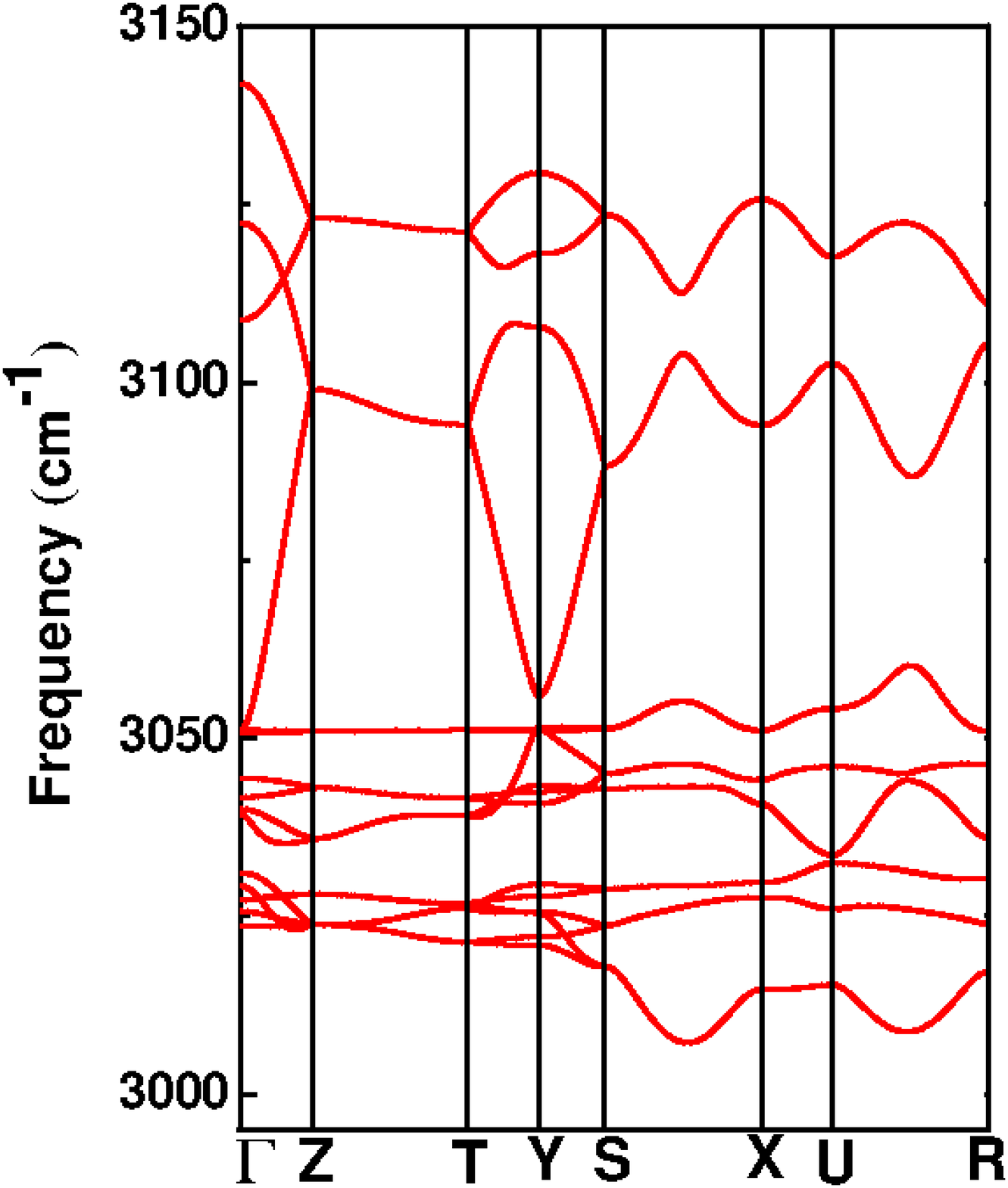}}}
{\subfigure[]{\includegraphics[height = 2.5in, width=3.0in]{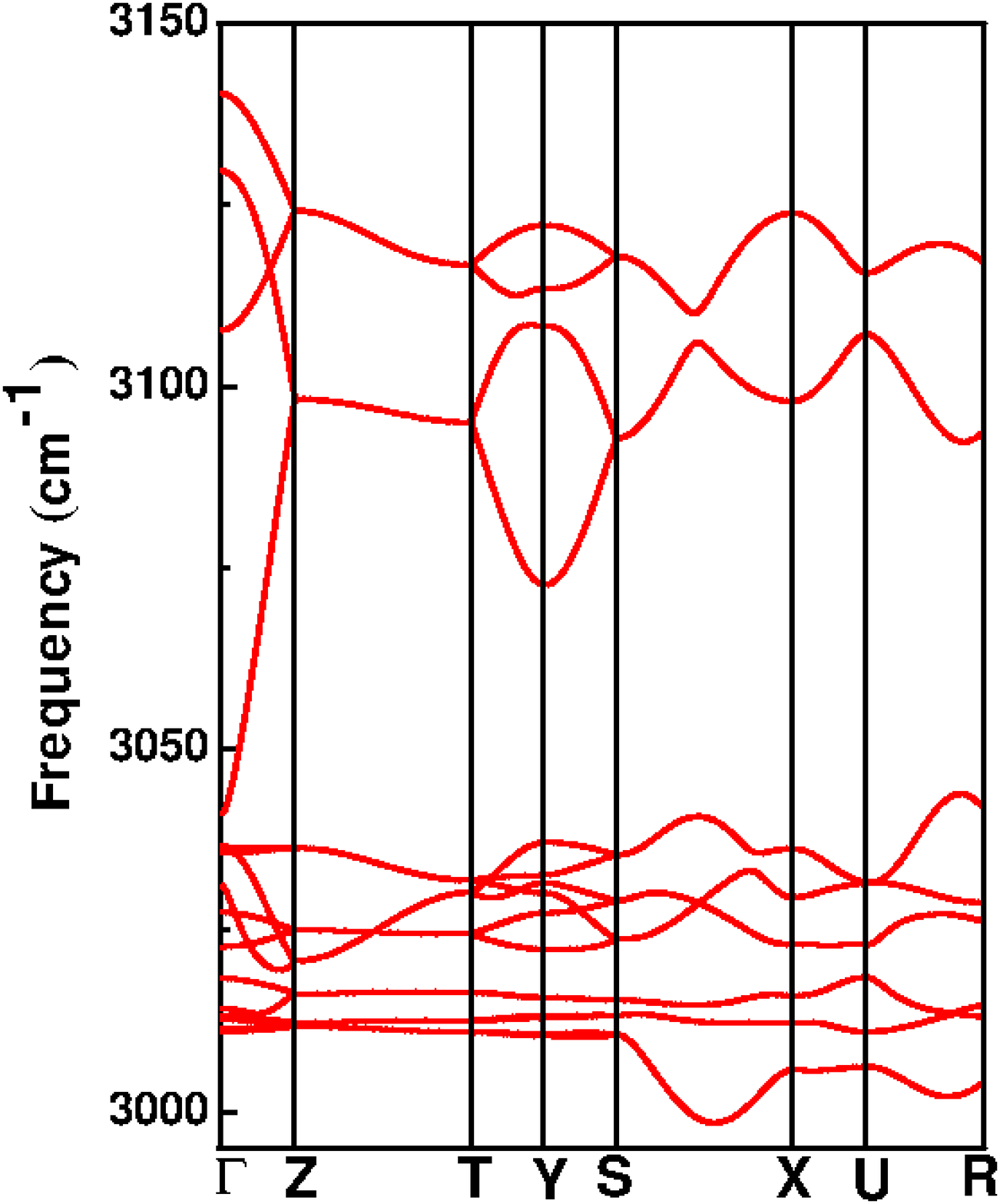}}}
\caption{(Colour online) Calculated phonon dispersion curves in the high frequency region ($\textgreater$ 700 cm$^{-1}$) of AA at 0 GPa (left) as well as at 6 GPa (right) along high symmetry directions (notations have same representation as in figure \ref{fig:disp_1}) in the Brillouin zone.}
\label{fig:disp_2}
\end{figure}

\clearpage
\begin{table}[h]
\caption{Calculated and experimental ground state lattice parameters (a, b, c, in $\AA$), crystallographic angles ($\alpha$, $\beta$, $\gamma$, in $^{\circ}$), volume (V, in $\AA^3$), and total energy (E$_0$, in eV/atom) of N$_4$H$_4$ compounds. USPP and NCPP were used in the vdW-TS calculations implemented in CASTEP package and PAW pseudo potentials were used within dispersion corrected TS-SCS method incorporated through VASP package.}
\begin{ruledtabular}
\begin{tabular}{ccccccc}
Phase            &  Parameter  &      &  This work  &        &     Other calculations      &  Experiment \\
                 &             &  USPP  &   NCPP   &  PAW    &                             &   \\ \hline
AA               &      a      & 9.015  &  9.012   & 9.021   &     9.0247$^a$, 8.7539$^b$  &  8.93283$^f$, 8.948$^g$, 8.937$^h$   \\
(Pmna, Z=4)      &      b      & 3.809  &  3.816   & 3.837   &     3.8089$^a$, 3.6363$^b$  &  3.80848$^f$, 3.808$^g$, 3.807$^h$   \\
                 &      c      & 8.551  &  8.574   & 8.482   &     8.5460$^a$, 8.3560$^b$  &  8.66147$^f$, 8.659$^g$, 8.664$^h$   \\
                 &      V      & 293.66 &  294.89  & 293.57  &      275.9$^c$, 285.4$^c$   &  294.45$^f$, 295.05$^g$, 294.79$^h$  \\
                 &      E$_0$  &-143.51 & -142.75  & -5.99   &    -143.512$^a$             &   \\
TTZ              &      a      & 10.469 & 10.498   & 10.391  &     10.5752$^a$             &  10.23$^i$   \\
(P$\bar{1}$, Z=4)&      b      & 6.888  & 6.904    & 7.030   &    6.8631$^a$               &  7.12$^i$   \\
                 &      c      & 4.081  & 4.127    & 4.122   &    4.0264$^a$               &  4.19$^i$  \\
                 &   $\alpha$  & 102.63 & 102.37   & 102.27  &   102.599$^a$               & 102.0$^i$  \\
                 &   $\beta$   &   87.94 & 88.19   & 88.16   &   87.463$^a$                & 90.0$^i$   \\
                 &   $\gamma$  &  104.22 & 103.79  & 104.89  &   104.248$^a$               &  106.5$^i$ \\
                 &      V      &  278.29 & 283.72  & 284.27  &                             &  285.66$^i$  \\
                 &      E$_0$  & -143.39 & -142.60 & -5.83   &   -143.392$^a$              &   \\
HNS-1            &      a      &  8.395  & 8.472   & 8.307   &  8.1850$^a$, 8.5935$^d$     &  \\
(P2$_1$/m, Z=2)  &      b      &  2.312  & 2.339   & 2.306   &  2.3114$^a$, 2.3167$^d$     &  \\
                 &      c      &  3.051  & 3.076   & 3.054   & 3.0641$^a$, 3.0021$^d$      &  \\
                 &   $\beta$   &  107.34 & 107.46  & 107.24  & 107.38$^a$, 105.74$^d$      &  \\
                 &      V      &  56.53  & 58.15   & 55.88   &                             &  \\
                 &    E$_0$    & -143.19 & -142.37 & -5.58   &  -143.158$^a$               &   \\
HNS-2            &      a      &  8.183  & 8.209   & 8.204   &  8.2348$^e$                 &   \\
(P4$_2$/n, Z=4)  &      c      &  4.211  & 4.367   & 4.290   &  4.1338$^e$                 &   \\
                 &      V      &  281.96 & 294.28  & 288.77  &                             &  \\
                 &    E$_0$    & -143.16 & -142.39 & -5.60   &  -143.191$^e$               &   \\

\end{tabular}
\end{ruledtabular}
\label{tab:lp}
$^a$ Ref.\cite{Liu1};
$^b$ Ref.\cite{Liu2};
$^c$ Ref.\cite{Landerville2};
$^d$ Ref.\cite{hu};
$^e$ Ref.\cite{Liu3};
$^f$ Ref.\cite{wu};
$^g$ Ref.\cite{prince};
$^h$ Ref.\cite{amorim};
$^i$ Ref.\cite{Veith}
\end{table}


\end{document}